\documentclass[useAMS,usenatbib,usegraphicx]{mn2e}

\usepackage{amsmath,paralist,xcolor,xspace}
\usepackage{units} 
\usepackage[normalem]{ulem}
\IfFileExists{srcltx.sty}{\usepackage[active]{srcltx}}

\newcommand{\msun}{M_\odot}
\newcommand{\schr}{Schr\"odinger\xspace}
\newcommand{\dd}[2]{\frac{\partial #1}{\partial #2}}
\newcommand{\lcdm}{$\Lambda$CDM }
\newcommand{\nbody}{$N$-body }

\newcommand{\sym}{\mbox{sym}}
\newcommand{\psiP}[1]{\psi_n^*\left(x+\frac{#1}{2}\right)}
\newcommand{\psiM}[1]{\psi_n\left(x-\frac{#1}{2}\right)}

\title[A framework for simulations of structure formation]{A new framework for numerical
simulations of structure formation}

\author[M. Schaller et al.]  {Matthieu Schaller$^1$\thanks{E-mail: matthieu.schaller@durham.ac.uk},
			      Claude Becker$^2$, 
			      Oleg Ruchayskiy$^2$,
			      Alexey Boyarsky$^{3,4}$ and\newauthor 
			      Mikhail Shaposhnikov$^2$\\
$^1$Institute for Computational Cosmology, Durham University, South Road, Durham, UK, DH1 3LE\\
$^2$Institut de Th\'eorie des Ph\'enom\`enes Physiques, \'Ecole Polytechnique F\'ed\'erale de Lausanne, CH-1015 
Lausanne, Switzerland\\
$^3$Instituut-Lorentz for Theoretical Physics, Universiteit Leiden, Niels Bohrweg 2, Leiden, The Netherlands\\
$^4$Bogolyubov Institute of Theoretical Physics, Kyiv, Ukraine
}

\begin{document}

\date{Accepted 2014 May 29. Received 2014 May 21; in original form 2013 October 27}

\pagerange{\pageref{firstpage}--\pageref{lastpage}} \pubyear{2014}

\maketitle

\label{firstpage}

\begin{abstract}
  The diversity of structures in the Universe (from the smallest
  galaxies to the largest superclusters) has formed under the pull of gravity from the tiny 
primordial   perturbations that we see imprinted in the cosmic microwave
  background. A quantitative description of this process would require
  description of motion of zillions of dark matter particles. This impossible
  task is usually circumvented by \emph{coarse-graining} the problem: one
  either considers a Newtonian dynamics of ``particles'' with macroscopically
  large masses \emph{or} approximates the dark matter distribution with a
  continuous density field. There is no closed system of equations for the
  evolution of the matter density field alone and instead it should still be
  discretized at each timestep.  In this work we describe a method of solving
  the full 6-dimensional Vlasov-Poisson equation via a system of auxiliary
  Schr\"odinger-like equations. The complexity of the problem gets shifted
  into the choice of the number and shape of the initial wavefunctions that
  should only be specified at the beginning of the computation (we stress that
  these wavefunctions have nothing to do with quantum nature of the actual
  dark matter particles). We discuss different prescriptions to generate the
  initial wave functions from the initial conditions and demonstrate the
  validity of the technique on two simple test cases.  This new simulation
  algorithm can in principle be used on an arbitrary distribution function,
  enabling the simulation of warm and hot dark matter structure formation
  scenarios.
\end{abstract}


\section{Introduction}
\label{sec:introduction}

The Lambda Cold Dark Matter ($\Lambda$CDM) cosmological model is the current
theoretical framework to describe the formation and evolution of large scale
structures in the Universe.  In this model, the growth of structures occurs
through the hierarchical collapse of a collisionless fluid of cold dark matter
(CDM).  Small initial perturbations grow through merging to create more and
more massive halos and complex sub-structures
(e.g. \citealt{Davis1985,Bertschinger1998, Springel2005}).  These initial
perturbations are thought to be (almost) Gaussian, created from quantum
fluctuations during the inflation epoch and are the origin of all the objects
seen in the Universe. The knowledge of the precise initial conditions and a
comprehensive understanding of the underlying physical laws should, in
principle, enable us to evolve these fluctuations forward in time and provide
a test of the current models.

Most of the important features observable in the Universe today have grown via
non-linear evolution from tiny primordial density perturbations. This makes
the whole process of understanding their evolution complex and requires the
use of techniques well beyond the linear perturbation theory
\citep{Bernardeau2002}. Indeed, at scales below roughly $\unit[10]{Mpc}$ the
evolution of structures had already entered the non-linear stage (i.e.\ the
density contrast $\delta \rho$ is of order (or much greater) than the
background density $\bar\rho$).  The main resource available to cosmologists
is the use of bigger and bigger cosmological simulations, most of them using
the particle technique known as \nbody simulation \citep{Hockney1988,
  Dehnen2011}.  Numerical simulations may, for instance, help shed some light
on the unknown nature of dark matter.

Clearly, the number of dark matter particles is way too large to track
individually each of them on a computer. Therefore most of the cosmological \nbody\
simulations use macroscopically large simulation ``particles'' (with their
masses ranging from masses much larger than DM particles up to the size of a small 
galaxy, $10^8-10^9 \msun$).

The problem of dark matter evolution in the Universe can be formulated as an evolution of a collisionless 
self-gravitating fluid.

The main tool used to describe this dark matter fluid is the \emph{phase space
density distribution} $f(x,v,t)$, defined such that
$f(x,v,t)dqdv$ represents the mass of material at position $x$ moving at velocity $v$ at
time $t$. This function is usually normalized such that its integral over all positions and velocities gives the total 
mass

\begin{equation}
 \int d^3x\int d^3v f(x,v,t) = M_{\rmn{tot}}.
\end{equation}
Notice that one could also normalize this integral to one or to the total number of particles in the system. When
integrating over velocity space only, one gets the usual mass density $\rho(x)$, whereas integrating over all space
returns the velocity distribution $d_v(v)$:

\begin{equation}
 \rho(x) = \int  f(x,v,t)d^3v, \qquad d_v(v) = \int  f(x,v,t) d^3x.
\end{equation}
This distribution function obeys the Liouville theorem \citep{Binney2008} and if the only force acting on the particle 
is the gravitational potential $U(x)$, we can write a closed system of equations for the formation 
of structures
\citep{Bertschinger1995, Bernardeau2002}:

\begin{equation}
  \begin{aligned}
    \dd{f}{\tau} + \frac{v}{a(\tau)}\dd{f}{x} - a(\tau)\nabla U \dd{f}{v} &= 0,  \\
    \nabla^2 U &= 4\pi Ga^2(\tau) \delta\rho,
  \end{aligned}
\label{eq:VP}
\end{equation}
where $x$ and $v$ are comoving coordinates and velocities, $a(\tau)$ is the scale factor and $\tau$ is the 
conformal time (We will
use this convention throughout this paper). This Vlasov-Poisson system has no
solution in the general case and the only way to handle it is to use numerical
techniques.  For completeness, we also give the expressions for the density
and density contrast:

\begin{eqnarray}
 \rho(x,\tau) &=& \frac{1}{a^3(\tau)}\int f(x,v,\tau)d^3v, \\
 \delta\rho(x,\tau) &=& \frac{1}{a^3(\tau)}\left(\int f(x,v,\tau)d^3v - \frac{M_{\rmn{tot}}}{V_{\rmn{tot}}} \right),
\end{eqnarray}
where $V_{\rmn{tot}}$ is the total comoving volume over which we average.

\section{Structure formation simulations}
\label{sec:theory}

The numerical analysis of the Vlasov-Poisson system of equations (\ref{eq:VP}) is very
challenging. The first reason is that the system is six-dimensional. Recent
simulations can only handle up to $64$ resolution elements in each spatial and
velocity space direction \citep{Yoshikawa2013} due to memory
restrictions. Even the use of the biggest supercomputers would not allow to go much beyond this figure. 

The second shortcoming of such technique is the development of fine-grained
structures that are very difficult to follow numerically. These become very
important in structure formation scenarios as clusters typically present many
matter streams and shell crossings.

Those two main shortcomings make the search for more advanced numerical scheme
important. The problem of
high-dimensionality could be removed if there were a way to use the density
field $\rho(x)$ instead of the probability distribution function
$f(x,v)$. This can be done by integrating the first few moments of the Vlasov
equation and then use techniques known for hydrodynamical simulations (see
e.g.~\citealt{Hockney1988}).  This technique is limited by the formal need to
integrate all moments and not just the first few ones to obtain an exact
solution. Instead of a 6D space, there is now a (formally) infinite number of
variables obeying an infinite series of equations.  \cite{Peebles1987}, for
instance, truncates the series and uses the first two moments (mass conservation,
Euler equation) of the collisionless Boltzmann equation to evolve in time the
initial perturbations. The framework reaches its limits whenever the velocity
dispersion of the fluid becomes important or when shell crossing occurs.


\subsection{\nbody simulations}
\label{ssec:simulations}

The other option to solve the system of equations (\ref{eq:VP}) is to use a particle method in which the 
distribution function is
sampled by a finite number $N$ of particles such that

\begin{equation}
 f(x,v) \cong \frac{1}{N} \sum_{i=1}^{N} m_i \delta\left(x - x_i\right)\delta\left(v - v_i\right).
\label{eq:nbody}
\end{equation}
Each particle or body is then evolved according to Newton's law under the influence of the gravitational potential
created by all the others as described by Poisson's equation. In other words, \nbody simulations solve the Vlasov 
equation via its characteristics by sampling the initial phase space distribution with a discrete number of particles. 
The number of
bodies is typically chosen as large as computationally feasible. The \nbody formalism is thus a Monte-Carlo
approximation of the Vlasov-Poisson system. The advent of large supercomputers combined with the development of more
efficient numerical algorithms has enabled the field of cosmological simulations to make considerable progress over the
last decades. Simulations such as the \emph{Millennium run} \citep{Springel2005} or \emph{Bolshoi simulation}
\citep{Klypin2011} are able to follow as many as a few billion particles.

The complicated part of the \nbody simulation is the evaluation of the forces
between pairs of particles. Over the years, many ingenious techniques (see
\cite{Dehnen2011} for a review) have been invented to reduce the algorithms
complexity for the force integration to $\mathcal{O}(N\log N)$ or even better
\citep{Dehnen2000}. All these techniques (tree-code, particle-mesh, P$^3$M, AMR, tree-PM,...)  do
however rely on particles and do, hence, share the same initial assumptions leading to the 
two following challenges. 

Firstly, since the dark matter fluid is supposed to be collisionless, one has to manually suppress artificial 
two-body collisions arising between the pseudo-particles introduced to sample the phase space distribution. This
is usually done by introducing an ad-hoc softening length and suppressing the gravitational force at scales below it
\citep{Dehnen2001}. \nbody simulations are run under the assumption that for a suitable choice of the smoothing, the
evolution of the $N$ pseudo-particles under the softened force should be the same as the gravitational evolution of the 
elementary dark matter particles.

The second challenge is to relate the particle distribution to the theoretical Vlasov-Poisson the particles are 
supposed to model. Despite its obvious relevance, it seems that the question of the precise quantitative importance of 
the discretization (\ref{eq:nbody}) and its effects is still not settled \citep{Joyce2008}.

As a matter of fact, there are no alternative
tools to study the cosmic structure formation with the same resolution as
\nbody simulations. This is of course not a limitation of the \nbody method
itself, but makes it more complicated to evaluate the possible errors of
\nbody simulations quantitatively, as there are basically no independent
results to compare with. For instance \citep{Ludlow2011} find a non-negligible
fraction of halos in CDM simulations that cannot be matched to peaks in the
initial density distribution and are possible artefacts of the \nbody
method. The different techniques used to calculate the forces are, of course, different and can lead to marginally 
different results for the same initial sampling of the field when the resolution limit is reached. They do, however,
all share the decomposition of $f(x,p)$ in a set of $N$ macroscopic particles and will, hence, share the consequences of
this Ansatz.

Spurious effects due to the discretization become more apparent when looking
at simulations of warm dark matter (WDM) or hot dark matter (HDM) cosmologies. The initial
matter power-spectrum entering such simulations is truncated below a certain
free-streaming scale related to the dark matter particle rest mass. Those
particles having a small mass, they also have a finite velocity distribution
function at every point in space, making the problem effectively 6
dimensional.  In practice, these velocities are neglected and the DM fluid is
treated in the cold fluid limit. These simulations are run using the same
\nbody framework but with an initial density and velocity power spectrum
truncated below the scale of interest. This should lead to a suppression of
small halos below a characteristic mass and the simulations ought to be able
to reproduce all structures with a mass above this limit. They could thus
quickly converge towards a solution.  \cite{Colin2000, Wang2007, Colin2008}
did, however, demonstrate that this is not the case and that spurious halos
form and merge to form structures below the theoretical mass
threshold. Various techniques are used in the literature to cure this
problem. \cite{Lovell2012}, for instance, filter their halo catalogues during
the post-processing of their simulations. The end results are thus free from
spurious halos but it does not solve the intrinsic discreteness problem
of the \nbody technique.

More details about these challenges and a comprehensive review of the topic can be found in \cite{Dehnen2011}.
Notice that this formalism is still a very active and lively area of research with alternative more advanced
formulations being proposed frequently. Some authors \citep{Abell2012, Shandarin2012} proposed recently to use 
tessellations of the 3D matter sheet in 6D space to track some of the phase space information. This may allow them to 
solve the coarse graining problem and reduce the impact of non-physical two body relaxations between the macroscopical 
particles. This formalism has lead to promising results in the study of WDM cosmology and the differences between the 
CDM and WDM halo mass functions \citep{Angulo2013}.

All the potential shortcomings of the \nbody formalism and the difficulty to evaluate their impact on the simulation
results make it important to develop another framework not based on a particle approach.

\subsection{An alternative framework}

Our framework resembles the attempt by \citep{Peebles1987} to use only the density field 
$\rho(x)$ and potential $U(x)$.
The main problem of such an approach is that there is no closed system of equations that includes only the density and 
gravitational potential.

The situation is different when looking at quantum physics. In this realm, all the phase-space information can be
encoded in a single function, the wavefunction $\psi(x)$ which does not depend on the velocity $v$. It is thus 
possible
to write a closed Schr\"odinger-Poisson system that would replace the Vlasov-Poisson one and that would only depend on
the spatial variable $x$ (See also \cite{Short2006} for a similar idea). This would effectively be a 3D system 
of equations but would allow to simulate the full 6D phase space and hence allow simulation of alternative cosmologies, 
such as the one including WDM or free-streaming neutrino contributions.
The principal difficulty is then to find a good mapping between the distribution function $f(x,v)$ of interest and its
``quantum'' equivalent $\psi(x)$ and vice-versa. This is achieved by using the so-called \emph{Wigner distribution
function}

\begin{equation}
f(x,v) \simeq  \int e^{\frac{i}{\hbar}vy}\psi^*\left(x+\frac{y}{2}\right)\psi\left(x-\frac{y}{2}\right)
d^3y.
\end{equation}

which obeys an equation similar to the Vlasov equation but is constructed from wave functions. The main
feature of this mapping is that the density field can simply be expressed as

\begin{equation}
\label{eq:1}
  \rho(x) = |\psi(x)|^2.
\end{equation}
However, the limitation of this approach is that one single wavefunction is in general not sufficient to encode all the
complexity of the distribution function and we would then use the more general version:

\begin{equation}
\label{eq:2}
\rho(x) = \sum_n |\psi_n(x)|^2.
\end{equation}
The summation index $n$ can, as a first thought, be understood as a sum over the  velocities $v$ that appear in
the distribution function $f(x,v)$. We somehow trade a 6D function for a (finite) set of 3D (complex valued) functions.
We will, however, demonstrate that the number of wavefunctions required can be very low (of order unity in some cases),
making the whole framework effectively 3D. 

It is important to stress from the onset that we are not trying to solve the evolution of structure formation at the
quantum level. Although we make use of quantum mechanics concepts, we merely use it as mathematical ``trick'' to
solve the Vlasov-Poisson system (\ref{eq:VP}). For this reason, the constant $\hbar$ appearing in our equations has to
be understood as a computational parameter whose value bares no relation to the actual Planck constant
$\hbar_{\rmn{phys}}= 1.0545 \cdot10^{-34}~\rmn{m}^2~\rmn{kg}~\rmn{s}^{-1}$.

Once the wavefunctions are built, they are evolved forward in time using \schr's equation. The density sourcing
Poisson's equation is obtained through equation (\ref{eq:2}) and one can then solve for the potential at each time step 
using standard techniques. This potential enters the \schr equation, closing the loop. We have, hence, built a closed 
system of equations using only a set of wave functions (which serve as a proxy for density) and the gravitational 
potential. 
\begin{equation}
\label{eq:SP}
\begin{array}{rcl}
 i\hbar \partial_t \psi_n &=& - \displaystyle\frac{\hbar^2}{2} \nabla^2_x \psi_n +
U\psi_n, \Big.\\
\nabla^2_x U &=& 4\pi G\delta\rho. \Big.
\end{array}
\end{equation}

The study of structure formation then becomes an exercise in solving $n$ copies of the \schr equation on a computer, 
which is a well-studied problem. The velocity distribution can be recovered by Fourier transforming the wave functions 
and if one is interested in the phase space distribution, one can apply the Wigner transform. This is, however, not part 
of the algorithm itself. This can be done in post-processing if necessary. The entire evolution of the system can be 
done at the ``quantum level'', i.e. using the wave functions alone. 

We stress that this is another approximation of the true underlying physical problem (equation \ref{eq:VP}) and that 
this framework, as any other, will have limitations. Some of these limitations and their relevance to the case of 
structure formation studies will be discussed in this paper. We will address those in the context of the science we are 
interested in and demonstrate how alternative cosmologies, including non cold dark matter scenarios, could effectively 
be simulated. 

The development of this framework has been pioneered by \cite{Widrow1993} and \cite{Davies1997} with the important 
difference 
that these authors use a single wavefunction and another way to map the distribution function in the quantum world.
Their general procedure is very similar to ours: 
sample the wavefunction from the initial phase space distribution, evolve in time using the Schr\"odinger-Poisson
equations and recover the final phase space distribution from the
wavefunction. 

Note also that another possible route, where the Hartree equation is used instead of the Schr\"odinger equation, has 
been explored by \cite{Aschbacher2001} and \cite{Frohlich2010}.

The time evolution of the Schr\"odinger-Poisson system is done using an explicit finite-differences scheme 
for the wave function and a FFT algorithm to solve the Poisson equation. We try to improve upon their algorithm for the 
time evolution as will be described below. \cite{Widrow1993} have made several simulations using this Schr\"odinger 
method
obtaining results in agreement with usual \nbody simulations. They also claim that their method is computationally
comparable to \nbody simulations making it a promising tool for cosmological
purposes.

These authors choose to use one single wave function to represent
the distribution function. This has important consequences on the validity of equation (\ref{eq:1}). By using one single
wave function, the phase space distribution built from it can not be everywhere positive and the authors have thus to
add an additional Gaussian smoothing.  We alleviate this shortcoming by using more than one wave
function and a different transformation from wave functions to phase space distribution. 

Their choice of Gaussian smoothed density also led them to a simple technique to generate the initial wave function. They
use a set of particles sampling the phase space distribution function exactly as in the case of \nbody simulations. They
can then turn each particle into a Gaussian in phase space by smoothing it and use this set of wave packets as their
initial wave function. 

In our approach, we depart from this need of an initial \nbody sampling by
considering other techniques to generate the set of wave functions. By doing
so, we allow for a completely generic distribution function and should, in
principle, not experience the consequences of an \textit{a priori}
artificial Monte-Carlo sampling of $f(x,v)$.

The second feature of our framework is the replacement of the Poisson equation by a Klein-Gordon equation for the 
potential $U(x)$:
\begin{equation}
 -\frac{1}{c^2}\frac{\partial^2}{\partial t^2}U + \nabla^2_xU = 4\pi G\delta\rho,
\end{equation}

 where $c$ is the numerical speed of gravity. This scalar
gravity equation is, once again, purely a mathematical trick to reduce the complexity of the original system 
(\ref{eq:VP}) and
 not an attempt to modify Newton's gravity. Such a replacement 
makes the framework entirely local and
does not require complicated integration methods for the Poisson equation. The complexity of the scheme is then
formally reduced to $\mathcal{O}(M)$, where $M$ is the number of mesh points in real space used in the simulation. In
this respect our approach also differs from the original work by \cite{Widrow1993}, who stick to the classical
Poisson form of gravity.

We stress that this step is not formally necessary. The well-known techniques used to solve Poisson's
equation on a mesh (FFT, Gauss-Seidel relaxation, etc.) can also be used in our framework. This change of equation for
gravity does just make the computations slightly faster in the cases where our approximation is valid. However, in the 
case of cosmological simulations with vastly different scales interacting, it is unclear how the Klein-Gordon equation 
for gravity would behave and defaulting to standard mesh techniques might be required.

\section{The algorithm in brief}
\label{ssec:algorithm}

Here we present the main algorithm of our framework, decomposed in a few
simple steps. A formal derivation and a discussion of the convergence and
accuracy of the method will be presented in the next section.

\begin{asparadesc}
\item[Step 0:] Choose the parameters of your simulation. The precision and
  speed of the method is
  governed by three  parameters $\hbar$, $c$  and $N$. The algorithm of
  choosing them is the following:\\
  The parameters $c$ and $\hbar$ are linked to the time and space resolution ($\Delta x$ and $\Delta\tau$ respectively) 
of the simulation via the Courant 
condition:
  \begin{equation}
    \frac{\Delta x}{\Delta\tau} > c
  \end{equation}
  and the condition on the stability of the discretized \schr equation:
  \begin{equation}
    \frac{\Delta x^2}{\Delta\tau} > \hbar
  \end{equation}
  The number of wavefunctions $N$ is chosen depending on the number of relevant modes of the decomposition in 
wavefunctions of the initial distribution function. The optimal value of $N$ is problem dependent and is also influenced 
by the algorithm chosen to discretize the distribution function. The details of this procedure will be given in section 
\ref{sec:IC}. The precision of the original accuracy is also dictated by the choice of $\hbar$. The ``quantum'' nature 
of the formalism imposes limitations on the precision of the description of position and velocity at the same point 
following the equivalent of Heisenberg's uncertainty principle. 
  
\item[Step 1:] Take an initial phase-space distribution function of the matter
  fields $f(x,v)$ (in the case of cosmological simulations, it is expressed
  via the power spectrum $P(k)$). Decompose the distribution function in $N$
  complex-valued $\psi_n(x)$ such that
\begin{equation}
\label{eq:initial_distribution}
f(x,v) \approx \sum_{n=1}^N  \int 
e^{\frac{i}{\hbar}v\,y}\psi_n^*\left(x+\frac{y}{2}\right)\psi_n\left(x-\frac{y}{2}\right)
d^3y.
\end{equation}
The number $N$ of wavefunctions is chosen such as to minimize the error
introduced by the decomposition and will, in practice, be as big as
computationally feasible. Various ways to generate this initial set of
wavefunctions for a given $f(x,v)$ are presented in section~\ref{sec:IC}.

\medskip

\emph{At this stage the precision of the approximation $f(x,v)\to
  \{\psi_n(x)\}$ is controlled by two parameters, $\hbar$ and $N$.}

\item[Step 2:] The wavefunctions $\psi_n(x)$ are now evolved forward in time
  using the coupled \schr-Klein-Gordon system of equations
\begin{equation}
\begin{array}{rcl}
  i \hbar \partial_t \psi_n &=& - \displaystyle\frac{\hbar^2}{2} \nabla^2_x \psi_n +
  U\psi_n, \Big.\\
  -\displaystyle\frac{1}{c^2}\displaystyle\frac{\partial^2}{\partial t^2}U + \nabla^2_xU &=& 4\pi 
  G\left(\displaystyle\sum_{n=1}^N |\psi_n(x)|^2 - \bar\rho\right). \Big.
\end{array}
\end{equation}

The integration in time of the \schr-Klein-Gordon system can be done
explicitly using finite differences on a regular grid, as will be described
in section \ref{sec:numerics}.

\item[Step 3:] Controlling your simulation. As the simulation is running you
  should monitor the following quantities in order to see that the choice of
  the method does not introduce artefacts. The correction terms
\begin{equation}
\label{eq:correction_term}
\sum_{\shortstack{$\scriptstyle r\geq 3$\\$\scriptstyle r~\rm {odd}$}}
\frac{1}{r!}\left(\frac{\hbar}{2i}\right)^{r-1} \dd{^r}{x^r}V \dd{^r}{v^r}f(x,v)
\end{equation}

should be small when compared to the ones \big($v\dd{f}{x}$ and $\dd{U}{x}\dd{f}{v}$\big)  entering 
the Vlasov equation. Thanks to the $1/r!$ decrease and the smoothness of the gravitational potential $V$ in most cases 
of interest, computing the first term of this series is generally sufficient.
If this term grows above the value of the other terms in the Vlasov equation, then the approximation introduced in this 
paper is not valid any more. Reducing the value of $\hbar$ or increasing the number $N$ of wavefunctions used in the 
initial discretization 
will decrease the contribution of the correction terms but this will lead to a higher computational cost.
The correction terms as well as the terms entering the Vlasov equation are expensive to compute but need not be computed 
at each time step.

\item[Step 4:] Once the final time has been reached, the distribution function can be
recovered by computing the integral (\ref{eq:initial_distribution}) or if one
is only interested in the density field only, then equation (\ref{eq:2}) is
sufficient and straightforward to compute.
\end{asparadesc}

\section{Formal derivation}
\label{sec:framework}

In the previous section, we described the problem we were interested and the usual schemes used in the literature. We
also presented a brief description of the route we intend to follow in order to tackle the issues outlined.  In this
section, we describe the whole formulation in detail, derive its main equations and discuss its limits.
For completeness, we start with a review of a formulation of quantum mechanics and show how its main ingredient, the
\emph{Wigner Distribution Function}, will play the role of an approximate distribution function for our problem.
Readers interested only in the end results can jump directly to Section \ref{ssec:local}.

\subsection{Phase-space quantum mechanics}
\label{ssec:QM}

Quantum mechanics is usually presented as emerging from the Hamiltonian formulation 
of classical mechanics through canonical quantization (See for instance \cite{Sakurai}). In this procedure,
variables are promoted to Hermitian operators and the Poisson bracket is replaced by a commutator. Alternatively, one 
can also use Feynman's propagator and the path integral formalism to move from classical to quantum mechanics.

Alongside these well-known quantization procedures, there exist other equivalent formulations which try to emphasize
more clearly certain aspects. The Moyal (or phase-space) formulation is among those and tries to find a quantum
equivalent to the classical phase-space and distribution functions \citep{Ercolessi2007,
Hillery1984}. The quantization procedure tries to find a correspondence between classical
functions (called symbols) of the phase space variables and quantum operators in Hilbert space:

\begin{equation}
 \mbox{Operators in Hilbert space}\leftrightarrow \mbox{Phase space symbols}
\end{equation}
As the position and momentum operators do not commute, this mapping can not be unique. Different operator orderings
will be mapped to different phase-space symbols. Hermann Weyl proposed a systematic way to associate a quantum operator
to a classical distribution function, which is now referred to as \emph{Weyl quantization}. This complex procedure
will not be discussed further here but its inverse, the \emph{Wigner transform} will be useful for our formalism.
This transformation associates to every quantum operator $\hat{A}$ a real phase-space function $A(x,v)$:

\begin{equation}
 A(x,v) = \sym(\hat{A}) := \int e^{\frac{i}{\hbar}vy}\left\langle x-\frac{y}{2}
\right|\hat{A}\left|x+\frac{y}{2}\right\rangle dy,
\label{eq:Wigner_transform}
\end{equation}
where $\langle \cdot|\cdot\rangle$ is the usual Bra-ket notation for quantum states.
The transformation of a product of operators is given by

\begin{equation}
 \sym(\hat{A}\hat{B}) :=\sym(\hat{A}) \star \sym(\hat{B}),
\end{equation}
where the \emph{Moyal star product} $\star$ contains the quantum mixing of the operators. This
product of functions in phase space is defined as
\begin{equation}
 A(x,v)\star B(x,v) := A(x,v)~
e^{\frac{i\hbar}{2}\left(\overleftarrow{\partial_x}\overrightarrow{\partial_v} -
\overleftarrow{\partial_v}\overrightarrow{\partial_x} \right)}~B(x,v)
\end{equation}
and is a central element in this formulation of quantum mechanics. Defining the \emph{Moyal bracket} \citep{Moyal1949}
by

\begin{equation}
 \left\lbrace A, B\right\rbrace_M := A\star B - B \star A
\end{equation}
the commutator of operators is associated to the Moyal brackets of two symbols in the following way:

\begin{equation}
 \sym\left(\left[\hat{A},\hat{B}\right]\right)
=\left\lbrace\sym\left(\hat{A}\right),\sym\left(\hat{B}\right)\right\rbrace_M.
\end{equation}
The dynamical equation in this formulation can be written in a simple way using these brackets and reads

\begin{equation}
 i\hbar \partial_t f = \left\lbrace \hat{H}, f \right\rbrace_M,
\end{equation}
where $\hat{H}$ is the Hamiltonian of the system. The interesting property of this formulation of 
quantum mechanics is that in the semi-classical limit $\hbar\rightarrow 0$, the dynamical equation reduces to the 
classical equation of motion expressed in terms of Poisson brackets

\begin{equation}
 \partial_t f = \left\lbrace H,f\right\rbrace_P = H \left(\overleftarrow{\partial_x}\overrightarrow{\partial_v} -
\overleftarrow{\partial_v}\overrightarrow{\partial_x} \right) f.
\end{equation}
This illustrates how the algebraic structures of classical and quantum mechanics are related through the continuous
changing of the parameter $\hbar$. This is the reason why such an approach to quantum mechanics is known as
\emph{deformation quantization} \citep{Hirshfeld2002}.

Let's now stop this overview and move to the part of this formalism which will be useful for the construction of our
new simulation framework.

\subsection{Wigner distribution function}
\label{ssec:WDF}

The Wigner transform (equation \ref{eq:Wigner_transform}) maps a quantum operator $\hat{A}$ to a classical function in
phase space. Wigner used this to associate a real phase space function to a quantum system \citep{Wigner1932}, now
called the \emph{Wigner distribution function} (WDF). It is defined as the symbol in phase space associated to the
density operator $\hat\rho$:

\begin{equation}
 P_W(x,v) := \sym\left(\hat\rho\right) = \int e^{\frac{i}{\hbar}vy} \left\langle x-\frac{y}{2}
\right|\hat{\rho}\left|x+\frac{y}{2}\right\rangle d^3y.
\end{equation}
As usual, the density operator can be expressed as the combination of pure state wavefunctions $\psi_n$:

\begin{equation}
 \hat\rho = \sum_n \lambda_n \left|\psi_n\right\rangle\left\langle\psi_n\right|, \quad \lambda_n \geq 0, \quad \sum_n
\lambda_n = 1.
\end{equation}
For mixed states, the WDF is thus 

\begin{equation}
 P_W(x,v) = \int e^{\frac{i}{\hbar}vy}\sum_n \lambda_n\psi^*_n\left(x+\frac{y}{2}\right)\psi_n\left(x-\frac{y}{2}\right)
d^3y,
\label{eq:WDF}
\end{equation}
while for a pure state, it reads

\begin{equation}
 P_W(x,v) = \int e^{\frac{i}{\hbar}vy}\psi^*\left(x+\frac{y}{2}\right)\psi\left(x-\frac{y}{2}\right)
d^3y.
\end{equation}
To simplify the expressions, we will use the notation $x_\pm = x \pm \frac{y}{2}$ and $\psi_{n\pm} = \psi_n(x_\pm)$ in
what follows.
The WDF has many similarities to the classical distribution function: $P_W(x,v)$ is a real function, as can be seen by
taking the conjugate and performing the change of variable $y\rightarrow-y$. It is also normalized to $1$ in the
following sense

\begin{equation}
 \int d^3x\int \frac{d^3v}{(2\pi \hbar)^3} P_W(x,v) = 1.
\end{equation}
It has similar marginal distributions as can be seen by integrating over all velocities:
\begin{eqnarray} 
\int\frac{d^3v}{(2\pi \hbar)^3} P_W &=& \sum_n
\lambda_n\int
\delta^3\left(y\right)\psi^*_n\left(x_+\right)\psi_n\left(x_-\right)d^3y\nonumber  \\
&=& \sum_n \lambda_n \left| \psi_n\left(x\right) \right|^2,
\label{eq:density}
\end{eqnarray}
or over all space

\begin{eqnarray}
 \int d^3x P_W &=& \sum_n \lambda_n \int d^3x_-
 d^3x_+e^{\frac{i}{\hbar}vx_+}e^{-\frac{i}{\hbar}vx_-}\psi^*_{n+}\psi_{n-} \nonumber \\
  &=& \sum_n \lambda_n\left|\tilde\psi_n\left(\frac{p}{\hbar}\right) \right|^2.
\end{eqnarray}
In both cases the non-negative property of these marginal distributions is a property of the wavefunctions in quantum
mechanics. 

The Wigner distribution function does, however, have the peculiar property that it may assume negative values. For this
reason, it is called a \emph{quasi-probability distribution} and cannot be interpreted as a phase-space probability
density in the sense of classical mechanics. The non-positivity of the WDF can be seen by integrating over all
phase-space the product of two distributions built from different states $\psi$ and $\Phi$:

\begin{equation}
 \int dx \int dv~ P_W[\psi](x,v) P_W[\Phi](x,v)  \propto \left| \left\langle \psi |\Phi\right\rangle\right|^2.
\end{equation}
The right-hand side vanishes if the two states $\psi$, $\Phi$ are orthogonal which implies that the WDF cannot be
positive everywhere. According to the Hudson theorem \citep{Hudson1974}, the WDF of a pure state is point wise
non-negative if and only if the state is Gaussian.
If $\hat\rho$ is not a pure state, it can be represented as a convex combination of pure state operators, $\hat\rho =
\sum_n \lambda_n | \psi_n \rangle\langle \psi_n|$, in infinitely many ways. The WDF satisfies the so-called mixture
property \citep{Ballentine1998},  which is the requirement that the phase space distribution should depend only on the
density operator $\hat\rho$ and not on the particular way it is represented as a mixture of some set of pure states
$\left\lbrace \left|\psi_n\right\rangle\right\rbrace$.
To summarize, the Wigner distribution function has many properties similar to the classical phase space distribution.
Nevertheless it has been realized from the early days, that the concept of a joint probability at a phase space point is
limited in quantum mechanics because the Heisenberg uncertainty principle makes it impossible to simultaneously specify
the position and velocity of a particle. Therefore, the best one can hope to do is to define a function that has a
maximum of properties analogous to those of the classical distribution function. Many different variants of
distribution functions - Husimi, Kirkwood-Rihaczek, Glauber - have been studied over the decades, all with their own
advantages and shortcomings (See \cite{Lee1995} for a review). The WDF is despite its non-positivity considered to be a 
useful
calculational tool and finds applications in various domains outside of quantum physics, like signal processing or
optics \citep{Bastiaans1997}.

\cite{Widrow1993} use a Husimi distribution \citep{Husimi1940} to recover the phase space
information from the wavefunction. The Husimi distribution is essentially equal to the Wigner distribution with an
additional Gaussian smoothing of width $\eta$

\begin{equation}
 P_H(x,v) = \frac{1}{(2\pi\hbar)^3}\frac{1}{(\pi\eta^2)^{3/2}}\left|\int d^3y
e^{-\frac{(x-y)^2}{2\eta^2}-\frac{i}{\hbar}vy} \psi(y) \right|.
\end{equation}
Compared to the WDF it has the advantage of yielding a phase space distribution that is positive-definite at every
point. This comes at the price of the marginal distributions not being equal to the usual position and velocity
distributions, but rather Gaussian broadened versions of it

\begin{equation}
 \rho_H(x) = \frac{1}{(\pi\eta^2)^{3/2}}\int d^3 y e^{-\frac{(x-y)^2}{2\eta^2}}|\psi(y)|^2.
\end{equation}
Only in the limit $\eta\rightarrow0$ does it reduce to the usual probability distribution.
Similarly one can show that the other marginal distribution reduces to the standard velocity distribution only when
$\eta\rightarrow\infty$ This complementarity is of course related to Heisenberg's uncertainty principle. Note that it is
in principle this smoothed distribution  that enters Poisson equation instead of $|\psi(x)|^2$. Since this would
requiring an additional space integration at each time step, Widrow and Davies approximate it with
the usual distribution $|\psi(x)|^2$ in the Poisson equation. 

Actually, $P_H(x,p) \simeq f(x,p)$ only when averaged on scales $\Delta x \geq \eta$, $\Delta p \geq \frac{\hbar}{\eta}$
. Note that there is no a priori reason why the non-linear time evolution should yield an answer that is again,
in average, close to the real distribution function. Let us stress that we allow for several wavefunctions to have an
initial phase space representation that is arbitrary close to the classical distribution function at every point,
not only when averaged.

Let us recall that our goal is not to interpret the Wigner distribution function as a fully-fledged phase space
distribution, but rather as a convenient mathematical tool.

\subsection{Dynamical equation for the WDF}
\label{ssec:EOM}

We now want to derive the dynamical equation satisfied by the WDF.
A derivation starting from Liouville's equation for the density matrix can be found in \cite{Ballentine1998}. Another
possibility is to start by taking the time derivative of the Wigner distribution function and use the fact that the
wavefunctions satisfy Schr\"odinger's equation.

Suppose each of the wavefunctions satisfies Schr\"odinger equation

\begin{equation}
 i\hbar \partial_t \psi_n = -\frac{\hbar^2}{2}\nabla^2 \psi_n + V\psi_n,
\label{eq:schroedinger}
\end{equation}
then the time derivative of the WDF becomes

\begin{eqnarray}
\partial_t P_W =\int e^{\frac{i}{\hbar}vy} \sum_n \lambda_n \left[
-\frac{i\hbar}{2}\right. \left. \left(\nabla^2_+\psi^*_{n+}\psi_{n-} \right. \right. 
\nonumber \\
\left.- \psi^*_{n+}\nabla^2_-\psi_{n-} \right) - \left.\frac{1}{i\hbar}\left(V_+-V_-\right)\psi^*_{n+}\psi_{n-}\bigg.
\right] d^3y,
\end{eqnarray}
where, once again, the subscripts $+$,$-$ denote the dependence on $x_\pm = x \pm \frac{y}{2}$. The terms containing a
Laplacian can be rewritten in terms of spatial derivatives of $P_W$ only and the previous equation becomes

\begin{eqnarray}
 0 &=&\partial_t P_W + \vec{v}\cdot\vec{\nabla}_x P_W \nonumber \\
&& - \frac{1}{i\hbar}\int e^{\frac{i}{\hbar}py}\left(V_+-V_-\right)\sum_n\lambda_n\psi^*_{n+}\psi_{n-}~d^3y.
\end{eqnarray}
This is the dynamical equation for the WDF, that we will refer to as the \emph{Wigner equation}. This dynamical equation
depends on both $P_W$ and the wavefunctions which implies that we might have to define initial conditions for 
both.
Let's now demonstrate that one can get rid of the dependency on the $\psi_n$. Let's expand the potential in Taylor
series

\begin{equation}
 V(x_+) - V(x_-) = y \dd{}{x}V(x) + 2\sum_{\shortstack{$\scriptstyle r\geq 3$\\$\scriptstyle r~\rm {odd}$}}
\frac{1}{r!}\dd{^r}{x}V(x)\left(\frac{y}{2}\right)^2
\end{equation}
and use this result in the dynamical equation:
\begin{eqnarray}
0 &=& \dd{}{t} P_W + \vec{v}\cdot\dd{}{x} P_W - \dd{V}{x}\dd{}{v}P_W  \nonumber\\
  & & + \sum_{\shortstack{$\scriptstyle r\geq 3$\\$\scriptstyle r~\rm {odd}$}}
\frac{1}{r!}\left(\frac{\hbar}{2i}\right)^{r-1} \dd{^r}{x^r}V \dd{^r}{v^r}P_W.
\label{eq:Wigner}
\end{eqnarray}
One can notice that the first three terms correspond to the classical Vlasov equation.

 In three cases, the Wigner
equation exactly coincides with the classical Vlasov equation: for a free particle ($V=0$), for a uniform field
($V\propto x$)  and for a harmonic oscillator ($V \propto x^2$). In general, there are additional terms that can be
interpreted as quantum corrections\footnote{It may sound surprising that the equation for the harmonic oscillator
reduces exactly to the classical Vlasov equation, even though we know that the quantum mechanical treatment introduces
discrete energy levels. In this case the quantum information is encoded purely in the initial conditions.} or simply 
higher-order corrections. In any
other case, corrections in the form of a power series in $\hbar$ will appear and modify the dynamic.
Note that in this derivation, the only assumption made on the $\lambda_n$ is that they be 
constant. In principle any value is
acceptable and it can even be negative or complex. As we are not using these equations to solve a quantum mechanics 
problem, where $\lambda_n>0$, we can use this fact to create more general sets of wavefunctions to approximate a given 
$f(x,v)$.

Note that the mass $m$ does not appear in the \schr equation in the same way that it does not appear in the Vlasov 
system. This, once again, illustrates that we are not solving the quantum mechanics evolution of the individual DM 
particles but rather find an approximation of the DM fluid evolution equation.

Let us recap what we have derived so far. By inspecting the Moyal formulation of quantum mechanics, we found a 
quantity, the Wigner distribution function $P_W$. This quasi-probability density function obeys the Wigner equation, an 
equation similar to the Vlasov equation but with additional terms in the form of a power series in $\hbar$.

\subsection{Semi-classical limit}
\label{ssec:limit}

The Wigner equation (\ref{eq:Wigner}) reduces to the classical Vlasov equation in the limit 
$\hbar\rightarrow 0$. Even though the quantum correction is formally $\mathcal{O}\left(\hbar^2\right)$, the derivatives 
of $P_W$ could
generate additional inverse powers of $\hbar$, making the semi-classical limit more involved\footnote{This formulation
of
the statement is not fully satisfying, as the true semi-classical limit is also a statement about the properties of the
wavefunction, and not identical to sending $\hbar\rightarrow 0$ which is anyway a dimensional parameter.}.

The properties of the semi-classical limit depend of course on the potential $V(x)$. In this paragraph we present some
results concerning the case of interest to us, where the potential satisfies Poisson's equation. In particular,
different authors investigated the semi-classical limit of the Wigner-Poisson (W-P) system to the Vlasov-Poisson (V-P)
system for the Coulomb potential.

The mathematically rigorous classical limit from W-P to V-P has been solved first in 1993 independently by
\citep{Lions1993} and \citep{Markovitch1993}. Both references consider a so-called completely mixed
state; i.e. an infinite number of pure states with a strong additional constraint on the occupation probabilities:

\begin{equation}
 \mbox{Tr}\hat\rho^2 = \sum_n \lambda_n^2 \leq C\hbar^3,
\end{equation}

where $C$ is a constant. Under this assumption, the classical limit of the solution to the 3D W-P system converges to
the solution of the V-P system. Note that the Wigner distribution function can also have negative values, whereas the
semi-classical limit is a true, non-negative distribution function. In both references, this was overcome by using a
Gaussian-smoothed Wigner function.

The situation for a pure state is completely different \citep{Zhang2002}. According to these authors, it appears that a
density operator which has the above property that the trace of its square tends to zero with the third power of the
Planck constant seems to be closer to classical mechanics than a pure state. For a pure state in 1D, the semi-classical
limit is not unique: examples have been constructed where different regularization schemes give different limits
\citep{Majda1994}. The question whether there exists a selection principle to pick the correct classical solution has
also been investigated but is not yet settled \citep{Jin2008}. No proof of the semi-classical limit from W-P to V-P is
known for the pure state case in 2D or 3D. 

For more details the reader is referred to the original papers or the review \citep{Mauser2002}. See also
\citep{Frohlich2007} for an alternative approach to the semi-classical limit. 

Finally, let us stress once again, that we seek to use our knowledge of quantum mechanics to simplify the resolution of
the mathematical problem presented in the Introduction. We are not trying to describe the physics of
structure formation at the quantum level nor trying to find a wavefunction for the entire Universe.

\subsection{Local interaction framework}
\label{ssec:local}

In Newtonian gravity, much like in classical electrodynamics, each body moves in the potential generated by all the
others. As both forces are long-ranged, the total force acting on each of the $N$ particles will be given by the sum of
the contributions from all the other particles, no matter how far away. In gravitational \nbody problems, the $N$
sampling bodies also receive a contribution from all the other bodies and a naive algorithm would require
$\mathcal{O}(N^2)$ operations for the force calculation at each time step. But it is well-known that this long-ranged
interaction through the potential can be replaced by a purely local interaction with a gauge boson or a spin-zero boson.
In this approach, each particle only interacts locally with the bosonic field.

We propose to reformulate the cosmological Vlasov-Poisson problem system (\ref{eq:VP})

\begin{eqnarray}
\dd{f}{t} + \frac{v}{a} \dd{f}{x} - a \dd{U}{x}\dd{f}{v} &=&0, \nonumber\\
\nabla^2 U &=& 4\pi Ga^2 \delta\rho,
\end{eqnarray}
as a purely local problem. To achieve spatial locality, we shall trade the real-valued phase space
distribution function $f(x,v)$ for a finite set of complex-valued wavefunctions $\left\lbrace \psi_n(x)\right\rbrace$.
For this we shall assume that the classical distribution function can be approximated by the Wigner distribution
function of some auxiliary mixed states:

\begin{equation}
 f(x,v) \simeq P_W(x,v) = \int e^{\frac{i}{\hbar}vy}\sum_n \lambda_n \psi^*_n(x_+) \psi_n(x_-)d^3 y
\end{equation}
The details of how this approximation is to be understood, and how we construct in practice the set of wavefunctions
$\left\lbrace \psi_n(x)\right\rbrace$ for any given $f(x,v)$ will be discussed in Section \ref{sec:IC}. For the time
being, let us assume that we have determined a set of wavefunctions such that the above approximation holds. 

The dynamical evolution of the WDF is given by the quantum-corrected Vlasov equation (the Wigner equation
(\ref{eq:Wigner})), or equivalently, by the Schr\"odinger equation (\ref{eq:schroedinger}) of the wavefunctions
interacting in a self-consistent way with a potential obeying the Poisson equation. The cosmological Vlasov equation in
an expanding Universe and expressed using conformal time $\tau$ is very similar to the classical one, up to the
replacements

\begin{equation}
 v \mapsto \frac{v}{a(\tau)}, \qquad V \mapsto a(\tau)U.
\end{equation}
Therefore the Schr\"odinger-Poisson system in the expanding universe becomes

\begin{equation}
\label{eq:SP_cosmo}
\begin{array}{rcl}
 i\hbar \partial_\tau \psi_n &=& - \displaystyle\frac{\hbar^2}{2a} \nabla^2_x \psi_n +
aU\psi_n, \Big.\\
\nabla^2_x U &=& 4\pi Ga^2\delta\rho, \Big.
\end{array}
\end{equation}
where $\delta\rho$ is the cosmological density contrast. The mass density $\left[\rmn{kg}~\rmn{m}^{-3} \right]$ relates
to the wavefunctions $\left[\rmn{kg}^{1/2}~\rmn{m}^{-3/2} \right]$ by

\begin{equation}
 \rho = \frac{1}{a^3} \int \frac{d^3v}{(2\pi\hbar)^3}f(x,v) = \frac{1}{a^3} \sum_n \lambda_n \left| \psi_n\right|^2.
\end{equation}
The normalization is chosen such that the phase space density integrates to
the total mass

\begin{equation}
 \int d^3 x\int \frac{d^3v}{(2\pi\hbar)^3}f(x,v) = \int d^3 x \sum_n \lambda_n \left| \psi_n\right|^2 = M_{\rmn{tot}},
\end{equation}
implying for the background density

\begin{equation}
 \bar\rho = \langle\rho\rangle = \frac{1}{V_\rmn{tot}} \frac{1}{a^3}\int d^3 x \sum_n \lambda_n \left| \psi_n\right|^2 =
\frac{1}{a^3}\frac{M_\rmn{tot}}{V_\rmn{tot}},
\end{equation}
where $V_\rmn{tot}$ denotes the total comoving volume. Therefore the density contrast $\delta\rho$ reads

\begin{equation}
 \delta\rho = \frac{1}{a^3}\left(\sum_n \lambda_n \left| \psi_n\right|^2 - \frac{M_\rmn{tot}}{V_\rmn{tot}} \right).
\end{equation}

In the semi-classical limit ($\hbar\rightarrow 0$), the Schr\"odinger-Poisson system
(\ref{eq:SP_cosmo}) formally reduces to the original Vlasov-Poisson system describing gravitational
structure formation. 

Notice that the total mass is conserved by construction as the normalization of the wavefunctions is a constant of
motion of the Schr\"odinger equation.

So far, we achieved locality in the sense that our set of equations does not explicitly depend on the velocity
variable $v$. We traded our 6 dimensional phase-space density function for a (possibly infinite) set of complex-valued
functions that depend on the space coordinate $x$ only. The numerical complexity of the problem has thus been
drastically reduced as long as the number of wavefunctions remains small. Before addressing this 
question, let us go one step
further and discuss the second equation of our system (\ref{eq:SP_cosmo}). 

The Poisson equation is a non-local equation as the Laplacian operator couples the contributions from
the whole space. This can, however, be changed by replacing the Laplacian by a d'Alembertian operator. With this
change, the Poisson equation becomes a Klein-Gordon equation and our transformed cosmological problem now reads

\begin{equation}
\label{eq:S_KG}
\begin{array}{rcl}
 i\hbar \partial_\tau \psi_n &=& - \displaystyle\frac{\hbar^2}{2a} \nabla^2_x \psi_n +
aU\psi_n, \Big.\\
-\displaystyle\frac{1}{c^2}\partial^2_{\tau\tau}U + \nabla^2_x U &=& 4\pi Ga^2\delta\rho. \Big.
\end{array}
\end{equation}
This system is entirely local, meaning that it can be numerically evolved in time on a grid by summing contributions of
local sampling points only. If the contribution of the term $-\frac{1}{c^2}\partial^2_{\tau\tau}U$ becomes small, then
this system reduces to the Schr\"odinger-Poisson system discussed previously. This is in particular true in the
non-relativistic limit $c\rightarrow \infty$. It is important to understand that the speed $c$ does not necessarily have
to take the value of the physical speed of light (or of gravity) $c_{\rmn{phys}} = 299792458~\rmn{m}~\rmn{s}^{-1}$. It
must simply be understood as a parameter that we
can use to approach the physical problem we are interested in (equation \ref{eq:VP}). As for $\hbar$, we are free to
choose this
parameter in a way that is convenient for our simulations, as long as we remain in the non-relativistic limit, meaning
that the gravitational field $U$ propagates much faster than the matter fields $\psi_n$. 

Note, however, that using a non-infinite speed for the mediator of gravity in cosmological simulations may also be of
some physical interest as the Poisson equation is, formally, only a weak-field approximation of the underlying Einstein
equations from which a finite speed for the gravity emerges. Thus, modifying this parameter may also yield interesting
physical results. 

Let us summarize what we achieved so far. Using the formalism derived in the Sections \ref{ssec:QM} to
\ref{ssec:limit}, we have been able to construct a completely local system of equations (\ref{eq:S_KG}) which
in the non-relativistic classical limit $\hbar\rightarrow 0$, $c\rightarrow\infty$ reduces to the problem of
cosmological structure formation. The probability density function can be computed at any time using the definition of
the WDF (equation \ref{eq:WDF}) but we stress that this operation is in general not necessary as one is usually
interested in the evolution of the mass density (equation \ref{eq:density}) only. 

Let us finally say that replacing
the Poisson equation by a scalar field is not strictly necessary as the algorithmic complexity of the problem has
already been drastically reduced by the introduction of the WDF. Having a \schr-Poisson system to solve instead of
equation (\ref{eq:VP}) is more accurate than our final system (\ref{eq:S_KG}). It does, however, simplify a lot the
numerical algorithms in some cases and does not seem to impact heavily the results as long as the parameter $c$ is 
chosen wisely. The effect of this choice on the evolution of highly clustered matter fields found in the low redshift 
Universe has not, however, not been explored.

\subsection{Lagrangian formulation}
\label{ssec:lagrangian}

The system of equations (\ref{eq:S_KG}) can be derived from a Lagrangian density using the Euler-Lagrange equations. We
consider a real scalar field $U$ interacting with the complex scalar matter fields $\psi_n$. The Lagrangian for this
system reads

\begin{eqnarray}
\mathcal{L} &=& \frac{1}{2c^2}\dot U ^2 - \frac{1}{2}\left(\vec{\nabla}U\right)^2 + \zeta\bar\rho U \nonumber \\
 & & + \zeta \sum_n\lambda_n\left[\frac{i\hbar}{2}\left(\psi_n^*\dot\psi_n - \dot\psi_n^*\psi_n\right)  \right.
\nonumber\\
 & & \qquad\qquad\quad- \left.\frac{\hbar^2}{2a}\vec{\nabla}\psi_n^*\cdot\vec\nabla\psi_n - |\psi_n|^2U\right]. 
\label{eq:lagrangian}
\end{eqnarray}
The equations of motion are found to be

\begin{eqnarray}
 i\hbar \dot\psi_n &=& - \frac{\hbar^2}{2a}\nabla^2 \psi_n + U\psi_n, \\
  -\displaystyle\frac{1}{c^2}\ddot U + \nabla^2 U &=& \zeta \left(\sum_n \lambda_n\left|\psi_n\right|^2 -
\bar\rho\right),
\end{eqnarray}
which is the system we derived in the previous section if we set $\zeta =
\frac{4\pi G}{a}$. The Hamiltonian density corresponding to the Lagrangian (\ref{eq:lagrangian}) is given by

\begin{eqnarray}
\mathcal{H} &=& \frac{1}{2c^2}\dot U ^2  + \frac{1}{2}\left(\vec{\nabla}U\right)^2\nonumber \\
 & & + \zeta \sum_n\lambda_n \frac{\hbar^2}{2a}\left|\vec{\nabla}\psi_n\right|^2 \nonumber \\
 & & + \zeta \left(\sum_n \lambda_n\left|\psi_n\right|^2 - \bar\rho \right)U,
\label{eq:hamiltonian}
\end{eqnarray}
which has a positive definite kinetic energy term for the scalar potential, as expected from a well-behaved theory.

One can also decompose this Hamiltonian in its various energy components. Doing so allows us to control the impact of
the dynamic term for the field $U$ and consider it a valid approximation of the underlying Vlasov-Poisson problem when
its value is much lower than the other energy components. Together with the computation of the 
higher order terms of the Wigner equation (\ref{eq:Wigner}), this measure of the impact of $c\neq\infty$ gives us a 
measure of the approximations we made and can thus help us assess the validity of the outcome of our simulations.

\section{Generating Initial Conditions}
\label{sec:IC}

In the previous section, we showed how one can trade the Vlasov equation for the phase space distribution function for
Schr\"odinger's equation for the wavefunctions, as this allows for the introduction of a scalar field as the mediator
of the gravitational force. Of  course we do not require the wavefunctions to have any intrinsic physical 
interpretation.
We rather consider them, just like the WDF, as a mathematical tool and not as fundamental entities. Still we are faced
with the problem of how to determine a set of wavefunctions such that their WDF corresponds to the initial classical
phase space distribution.

One possible approach is to start from a set of $N$ particles sampling the phase-space distribution function and build
Gaussians centred on each point with a certain width $\eta$

\begin{equation}
 |\eta(x_i,v_i)\rangle\propto e^{-\frac{(x-x_i)^2}{2\eta^2}-\frac{i}{\hbar}v_i \cdot x}.
\end{equation}

The wavefunction is then obtained from the incoherent superposition of these wave-packets for each “particle” 

\begin{equation}
 |\psi\rangle = \frac{1}{\sqrt{N}}\sum_{i=1}^N e^{i\phi_i}|\eta(x_i,v_i)\rangle,
\end{equation}
where $e^{i\phi_i}$ is a random phase. This sampling procedure relies
on the assumption that each “particle” has a well-defined velocity. It is unclear how it could be generalized to the
case of warm dark matter, where the velocity dispersion is important. We remove the need for this
assumption by allowing for several wavefunctions. At the same time this allows us to represent any initial phase space
distribution without relying on \nbody sampling. Such an approach was used by Widrow \& Kaiser and is well-suited for
Husimi distributions as they contain an extra Gaussian smoothing. We will, instead, try to work directly with the
distribution function without sampling it in particles and hence taking the risk of facing the coarse graining and
discreteness effects (Section \ref{ssec:simulations}) that we are trying to avoid in our framework.

Since the wavefunctions encode both, the position and velocity information, a single
wavefunction (pure state) can in general not be sufficient to describe a generic $f(x,v)$. One should rather look for a
set of wavefunctions (mixed state). The more wavefunctions we allow for, the more freedom we have and the more
accurately the WDF should represent any given distribution. At the same time the total number of wavefunctions should
be
as small as possible because this will reduce the computational complexity of our numerical simulations.

Given the classical distribution function $f(x,v)$, we want to expand it using the WDF Ansatz

\begin{equation}
 f(x,v) = \sum_{n=1}^{N} \lambda_n \int e^{\frac{i}{\hbar}vy} \psiP{y} \psiM{y} d^3y.
\end{equation}
Fourier transforming from $v$-space to $\eta$-space we get

\begin{equation}
 f(x,\eta) = \sum_{n=1}^{N} \lambda_n \psiP{\eta} \psiM{\eta}.
 \label{eq:Fourier_WDF}
\end{equation}
Finding the wavefunctions is now a simpler problem provided one can easily compute the Fourier transform of the
distribution function one is interested in. We will discuss different approaches to tackle this problem of determining
 the set of wavefunctions $\psi_n$ and weights $\lambda_n$ representing a given initial phase space distribution
$f(x,v)$. Let us stress from the outset that these procedures need only to be used once at the beginning of a numerical
simulation, to set up the initial conditions. 

Last but not least, we need to emphasize that the number of wavefunctions is preserved by the quantum mechanical
evolution. There is no evolution equation for $\lambda_n$. Only the shape of the $\psi_n$ will change. This 
shows that it is the complexity of the initial conditions
that dictates the number of wavefunctions required. In a setup where only a restricted number of harmonics are present
in the initial probability distribution, already relatively few wavefunctions would be sufficient to represent
the system and its time evolution.

\subsection{Brute-force minimization}
\label{ssec:minimization}

The first and obvious method we present to choose the initial wavefunctions is a brute-force minimization. The
underlying idea is to define a functional measuring the total absolute error made by approximating the phase space
distribution by the WDF Ansatz

\begin{equation}
\Phi := \int d^3q \int d^3\eta \left|f(x,\eta) - \sum_{n=1}^{N}\lambda_n \psi_{n+}\psi_{n-}
\right|^2 ,
\end{equation}
where, once again, $\psi_{n\pm} = \psi_n\left(x\pm \frac{\eta}{2}\right)$. We
can then determine a set of wave functions that minimizes this error. In
practice, the minimization is most easily done via discretization on a
lattice. The problem is then cast into a minimization of the scalar error
function with a large number of variables corresponding to the values of the
wavefunctions at the lattice points. For different $N=1,2,\ldots$, we can
determine the set of wavefunctions $\psi_n$ and corresponding weights
$\lambda_n$ which minimizes the error. One can then compare the results for
different $N$ to find an optimal approximation with a high enough accuracy and
a minimal number of wavefunctions.

Since we are not seeking a true quantum mechanical interpretation, let us
consider the most general case of complex-valued weights. A naive minimization
will not yield wavefunctions normalized to unity. Instead of adding this
normalization as a constraint to the minimization, we remove the amplitude of
the complex weights $\lambda_n$, and only keep their phases $e^{i\phi_n}$.The
amplitudes of the weights are taken to be the norm of the wavefunctions,
thereby
normalizing them to unity. If we simply minimize the error functional, we will in general obtain
wavefunctions that are not smooth enough on the lattice to be evolved
numerically. For this purpose it is useful to add a term of the form of a kinetic term
 to the functional that
will allow us to enforce a certain degree of smoothness. We construct the
kinetic term from the square of the derivatives with a certain overall factor
$\chi$ to tune the smoothness:

\begin{equation}
 \mathcal{K} = \chi \int d^3x \sum_{n=1}^N \left|\dd{\psi_n}{x}\right|^2.
\end{equation}
Finally, we minimize this kinetic term with the total error summed over all lattice points

\begin{equation}
 \mathcal{E} = \mathcal{K} + \Phi.
\end{equation}
We have applied the method to cosmic initial conditions of cold dark matter 
in the Zel'dovich approximation, for
simplicity in a one-dimensional case. The results confirm the expectation that, increasing the number of wavefunctions,
the total error is reduced. In the case we studied, it turned out that already a relatively small number of
wavefunctions (compared for instance to the number of lattice points) was enough to achieve a reasonable accuracy.
As usual with minimization procedures, there is no guarantee that the algorithm converges to a global minimum. This
would for instance mean that one has to repeat the minimization with different initial random seeds and compare their
outcomes. Also, even though this minimization was shown to work for a given phase space distribution $f(x, v)$, in
practice it becomes computationally challenging even for rather small 3D lattice sizes, as the number of variables in
the minimization procedure grows quickly. Despite its applicability to any distribution function, the  brute-force
minimization might not be the best method to determine the initial wavefunctions.

\subsection{Eigenvalue problem for Hermitian operator}
\label{ssec:operator}

We now turn our attention to obtaining an analytic solution to the problem of determining the initial wavefunctions.
More precisely we will show how the Wigner Ansatz can be reformulated as an eigenvalue problem, which we can then solve
analytically in some specific cases.

Since $f(x,\eta)$ is the Fourier transform of a real function $f(x,v)$, it satisfies the condition
$f^*(x,-\eta)=f(x,\eta)$. Introducing the coordinates $x_\pm := x \pm \frac{\eta}{2}$, we can define
\begin{equation}
 g(x_-,x_+) := f\left(\frac{x_++x_-}{2}, x_+-x_-\right),
\end{equation}
which is then Hermitian 

\begin{equation}
 g^*(x_+,x_-) = g(x_-,x_+).
\end{equation}
Hilbert-Schmidt's theorem states that any square-integrable Hermitian kernel can be expressed in terms of its spectral
decomposition

\begin{equation}
 g(x_-,x_+) = \sum_n \lambda_n\psi^*_n(x_-)\psi_n(x_+),
\label{eq:spectral}
\end{equation}
where the $\lambda_n$ are real eigenvalues and ${\psi_n}$ the set of orthonormal eigenfunctions with respect to the
standard scalar product on $L^2(\mathbf{C}^3)$

\begin{equation}
 \left\langle \psi_n | \psi_m\right\rangle := \int \psi^*_n(x)\psi_m(x) d^3x = \delta_{nm}.
\label{eq:scalar}
\end{equation}
The Fourier space WDF (equation \ref{eq:Fourier_WDF}) has exactly the same form as the spectral decomposition
(equation \ref{eq:spectral}). Therefore we conclude that \emph{any} given phase space distribution function $f (x, v)$
can be written exactly as a WDF, if need be with an infinite number of wavefunctions. The wavefunctions are the
eigenfunction of the Hermitian operator $g(x_- , x_+ )$ and its real eigenvalues correspond to the weights of the
wavefunctions in the mixed state. Notice though, that they can in general take negative values, implying that we cannot
give a full quantum-mechanical interpretation to the mixed state, as the corresponding density operator is not
positive-definite. Let us emphasize once more that we consider the wavefunctions as a mere mathematical tool.

Multiplying both sides of (\ref{eq:spectral}) by $\psi_\alpha(x_-)$ and integrating over $x_-$ , the
orthonormality of the eigenfunctions implies the following integral equation

\begin{equation}
 \int g(x_-,x_+) \psi_\alpha(x_-) d^3x = \lambda_\alpha \psi_\alpha(x_+).
\end{equation}
This equation shows that the determination of the wavefunctions reduces to finding the eigenfunctions of the Hermitian
kernel  $g$. Unfortunately, for a completely general phase space distribution function, the above equation might
not allow an analytic solution. 

This procedure can be generalized by allowing for a more general scalar product containing a non-trivial weight function
$w(x)$:

\begin{equation}
 \left\langle \psi| \phi \right\rangle_w := \int \psi^*(x) \phi(x) w(x) d^3x.
\end{equation}
For such a scalar product, the eigenvalue decomposition of $g(x_-,x_+)$ still exists but the eigenfunctions are now
orthonormal with respect to the weighed scalar product. The eigenvalue problem thus reads

\begin{equation}
  \int g(x_-,x_+) \psi_\alpha(x_-)  w(x_-)d^3x = \lambda_\alpha \psi_\alpha(x_+).
  \label{eq:scalar_weight}
\end{equation}
Let us emphasize that the weighted scalar product is only used to determine the wavefunctions whose WDF equals the
classical distribution function. The choice of $w(x)$ is completely arbitrary and does not affect the properties of the
WDF or the Schr\"odinger evolution of the wavefunctions. Clearly the spectrum will depend on the choice of weight
function. The additional freedom of choosing $w(x)$ could allow to reduce the number of wavefunctions needed in the
Wigner Ansatz. Furthermore the arbitrariness of the weight function also reflects the freedom we have to
choose wavefunctions representing the initial state.

\subsection{Fourier-series decomposition}
\label{ssec:fourier}

Let us study the eigenvalue problem for a phase space distribution of the form\footnote{For the sake of simplicity we
restrict the analysis of this section to the one dimensional case, but the generalization to the 3D case
is straightforward.} $f(x, v) = \rho(x)\delta(v)$, meaning the product of a generic distribution in space with a delta
function in velocity space. This choice corresponds to the case of CDM at early times, when the velocities are 
negligible. 

In such a case, the integral operator $g(x_-,x_+)$ becomes real and symmetric

\begin{equation}
 g(x_-,x_+) = \rho\left(\frac{x_++x_-}{2} \right).
\end{equation}
We choose the trivial weight function $w(x) = 1$, which might not be the optimal choice for a minimal number of
wavefunctions, but yields a working example of the method. We now assume a periodic distribution of matter in $[0,L]$
and expand the density as a Fourier series over the interval 

\begin{equation}
 \rho(x) = \rho_0 + \sum_{n=1}^\infty  a_n \cos\left(\frac{2\pi n}{L}x \right) + \sum_{n=1}^\infty b_n
\sin\left(\frac{2\pi n}{L}x \right).
\end{equation}
The term $\rho_0$ can be dropped without loss of generality as it can trivially be represented in the WDF using a
constant wavefunction. 

The eigenvalue problem is easier to solve on the doubled interval $[0,2L]$. The standard scalar product for real
functions on this interval is simply

\begin{equation}
 \left\langle\psi|\phi \right\rangle := \frac{1}{L} \int_0^{2L} \phi(x) \psi(x) dx,
\end{equation}
which means that the eigenvalue problem reads

\begin{equation}
 \frac{1}{L} \int_0^{2L}  \rho\left(\frac{x+y}{2} \right) \psi(y) dy = \lambda \psi(x).
\end{equation}
We now have to choose an orthonormal basis for the wavefunctions $\psi$. As we work with a periodic interval, it is
natural to use harmonic functions over $[0,2L]$. The most general case is thus 

\begin{equation}
 \psi(x) = \sum_{n=1}^\infty \left[\alpha_n \cos\left(\frac{\pi n}{L}x \right) +  \beta_n \sin\left(\frac{\pi n}{L}x
\right)\right].
\end{equation}
Using trigonometric identities and the orthonormality relations between the sine and cosine functions of different
modes, the problem can be recast in a matrix problem for the coefficients of the Fourier series:

\begin{equation}
 \left(\begin{array}{cc}
        a_n & b_n \\
	b_n & -a_n
       \end{array}\right)
 \left(\begin{array}{c}
        \alpha_n \\
	\beta_n
       \end{array}\right)
 = \lambda
 \left(\begin{array}{c}
        \alpha_n \\
	\beta_n
       \end{array}\right).
\end{equation}
Therefore, the normalized eigenfunctions and eigenfunctions of the integral operator are finally given by

\begin{equation}
 \psi_n^\pm(x) = \mathcal{N}\left[ \left(a_n + \lambda_n^\pm\right) \cos\left(\frac{\pi n}{L}x \right)+ 
b_n \sin\left(\frac{\pi n}{L}x \right)\right], \label{eq:Fourier_decomposition}
\end{equation}
\begin{equation}
\lambda_n^\pm = \pm \sqrt{a^2_n + b^2_n},
\end{equation}
where $\mathcal{N} = \left[\left(a_n \pm \sqrt{a_n^2 + b_n^2} \right)^2 + b_n^2\right]^{-1/2}$ normalizes the
eigenfunctions to unity. It can be checked explicitly that these eigen-vectors satisfy the condition of orthonormality
and yield the correct spectral representation

\begin{eqnarray}
 \rho\left(\frac{x_+ + x_-}{2}\right) = \sum_{n=1}^{\infty} \left[\lambda_n^+ \psi_n^+(x_+)\psi_n^+(x_-)\right.
\nonumber \\
 \quad~+ \left.\lambda_n^- \psi_n^-(x_+)\psi_n^-(x_-)\right].
\end{eqnarray}
corresponding to the WDF

\begin{eqnarray}
 P_W(x,v) = \sum_{n=1}^{\infty} \int e^{\frac{i}{\hbar}vy}\left[\lambda_n^+ \psi_n^+\left(x + \frac{y}{2}
\right)\psi_n^+\left(x - \frac{y}{2}\right)\right.
\nonumber \\
 + \left.\lambda_n^- \psi_n^-\left(x+\frac{y}{2}\right)\psi_n^-\left(x-\frac{y}{2}\right)\right] d^3y.
\end{eqnarray}

As a conclusion we have been able to solve the eigenvalue problem on the finite interval and use it to find the
wavefunctions for the WDF Ansatz. This applies for a generic density profile $\rho(x)$ periodic on $[0,L]$ and a phase
space distribution of the form $f(x, v) = \rho(x)\delta(v)$. The wavefunctions are harmonic functions with increasing
velocity. In general we would need an infinite number of wavefunctions to avoid smoothing the smallest scales of 
the power spectrum. For many applications a finite or even small number of wavefunctions may be sufficient. 

In this procedure, we used the geometry of the problem to decide which orthonormal basis to use. The periodicity of the
density distribution naturally led us towards the use of harmonic functions. In cases were the density is not periodic,
one could use Chebyshev polynomials or any other basis whose geometry helps reduce the number of modes.

As already mentioned, the technique presented in this section holds for any power spectrum and in particular is well
suited to the case of WDM without initial velocities as is usually done in numerical simulations. This truncated CDM
power-spectrum can easily be decomposed in a Fourier series and hence used in our framework. If the thermal velocities
of the WDM particles have to be included, then another technique has to
be used (see sections \ref{ssec:operator} and \ref{ssec:SVD}).

\subsection{Cosmological initial conditions}
\label{ssec:cosmo_IC}

Observations of structure in the universe are perfectly compatible with the simplest possible statistical description,
namely a Gaussian distribution. More precisely, each Fourier mode of the density contrast $\delta(\vec{k})$ (not to be
confused with the Dirac delta distribution) satisfies an isotropic Gaussian distribution, entirely described by the
power spectrum $P(k):= \langle|\delta(\vec(k))|^2\rangle$, which is a function of the modulus $k$ only, not
of the direction. From the knowledge of the power spectrum one can then generate a realization with the desired
statistical properties 

\begin{eqnarray}
  \delta(\vec{x}) &=& \sum_{\vec{k}}\left[\sqrt{P(k)}\mathcal{N}(0,1)\cos(\vec{k}\cdot\vec{x})\right. \nonumber\\
 &&\left.\qquad+\sqrt{P(k)}\mathcal{N}(0,1)\sin(\vec{k}\cdot\vec{x})\right],
\end{eqnarray}
where $\mathcal{N}(0,1)$ denotes a Gaussian random number with zero mean and unit dispersion. This shows that the
density contrast for cosmological initial conditions is in a form for which we know how to construct the WDF, provided
that we start our simulation at times, when the Zel'dovich velocities of the particles are negligible. Compared to
\nbody simulations we do not need to first perform a FFT to compute $\delta(\vec{x})$ but can find the initial
wavefunctions directly from the power spectrum. Additionally we do not need any glassy pre-initial conditions to model
the constant background.

There is, however, a little caveat when generating initial conditions for CDM. Such an initial spectrum is formally
made of a Dirac distribution in $v$-space which means that even an infinite number of continuous wavefunction can not
reproduce exactly this singularity. This can also be explained by the quantum aspect of our formalism. Heisenberg's
uncertainty principle forbids us to have at the same time an infinitely precise description of position and velocity of
our wavefunction. There will be some necessary spread in velocity space proportional to the value of $\hbar$ chosen in
the simulation. The spectrum obtained will thus formally not exactly be the CDM one but will contain some intrinsic
velocities for the DM. These would vanish in the limit $\hbar\rightarrow0$.

We would in principle require as many wavefunctions as Fourier modes are relevant in the power spectrum, which may lead
to a prohibitive computational cost. Expanding the power-spectrum in an other basis or using a non-trivial weight
$w(x)$ in the scalar product (\ref{eq:scalar_weight}) may help reduce the number of wavefunctions required.
On the other hand, we may turn this as an advantage as this new formalism can allow us to probe some parts of the power
spectrum only without having to use the full range of $\vec{k}$.

\subsection{Matrix formulation}
\label{ssec:SVD}

Given that the WDF Ansatz can be thought of as spectral decomposition of an Hermitian operator, we can now analyse the
solution in the discrete case, where the problem reduces to a matrix problem. Let us again restrict the analysis to one
dimension. Working on a lattice $(x_1, x_2, \ldots,x_M)$, we can think of any function $f(x)$ as a vector $(f(x_1),
f(x_2), \ldots, f(x_M))^T$ and of any function of two variables as a matrix. We can thus reinterpret the functional
relationship

\begin{equation}
 g(x_-,x_+) = \sum_{n=1}^N \lambda_n\psi^*_n(x_-)\psi_n(x_+)
\end{equation}
in terms of matrices
\begin{equation}
 \hat G_{ij} = \sum_{n=1}^N \lambda_n {\Psi}_{jn}^*{\Psi}_{in} = \sum_{n=1}^N\sum_{k=1}^N
{\Psi}_{in}\lambda_n\delta_{nk}{\Psi}_{kj}^\dagger.
\end{equation}
The property $g(x_+,x_-) = g^*(x_-,x_+)$ translates into the fact that $\hat G \in\mathbf{C}^{M\times M}$ is a Hermitian
matrix $\hat G^\dagger = \hat G$ which we can diagonalize by means of a unitary transformation

\begin{equation}
 \hat G_{ij} = \left({\Psi}\cdot\hat{\Lambda}\cdot{\Psi}^\dagger\right)_{ij},
\end{equation}
where
\begin{equation}
 {\Psi}\in\mathbf{C}^{M\times N}, \qquad\hat{\Lambda} =
\mbox{diag}(\lambda_1,\ldots,\lambda_N)\in \mathbf{R}^{N\times N}.
\end{equation}
The columns of $\Psi$ are the wavefunctions $\psi_n$ sampled on the lattice. The property that $\Psi$ is
unitary $\Psi^\dagger\Psi = \mathbf{1}$ implies that the normalization of the wavefunctions on the lattice.
This matrix formulation has the advantage, that it is straightforward to compute the spectrum of any given Hermitian
matrix. The shortcomings of this approach are two-fold: firstly we would need as many wavefunctions as lattice points,
which comes at a big computational cost, and secondly the eigen-vectors have no a priori reason to be smooth enough to be
used as initial conditions for our numerical scheme. Note, however, that in all the cases we tested, the eigenvalue
decomposition has yield smooth enough functions. 

 Moreover it has to be noted that we would need to compute the eigen-vectors for a matrix containing the full 3D lattice.
Computing the eigen-vectors of a $n \times n$ matrix is in general a problem of complexity $\mathcal{O}(n^3)$. Since the
size of the matrix is related to the number of lattice points $M^3$, one quickly reaches such lattice sizes 
making the
solution of the eigenvalue problem impossible. This issue can be solved by combining this technique with the
minimization procedure. One can first use an eigenvalue decomposition on a coarse grid and use this as an input of the
brute-force minimization algorithm on a finer grid. A technique using multiple grids at the same time could also be
used in the same way that Gauss-Seidel relaxation is done in some particle-mesh gravity solvers. 

There are multiple known algorithms available to decompose a matrix in eigen-vectors. We chose to use the singular value
decomposition (SVD) as the publicly available implementations return the eigenvalues sorted in decreasing order. This
allows us to choose only the wavefunctions whose eigenvalues are above a certain (arbitrarily chosen) level.

\subsection{Discussion and remarks}
\label{ssec:remarks}

For numerical simulations in a finite box with periodic boundary conditions, the spatial lattice resolution also
dictates the resolution in velocity space. The size of the box is related to the lattice size in $v$-space since the
wave-vectors take discrete values $\vec{v} = \frac{2\pi}{L}\vec{n}$. The maximal wave-vector is linked to the lattice
spacing in real space. This illustrates the relationship between the number of wavefunctions and the spatial resolution
of the simulation. If we keep all the modes, we need $\mathcal{O}(M^3)$ wavefunctions, where $M$ is the
number of lattice points in one direction. Note that this corresponds, in order of magnitude, to the number of particles
in \nbody simulations. So even if we keep the maximal number of wavefunctions needed to accurately represent the
initial conditions, the complexity of our numerical scheme will still be comparable to the naive $\mathcal{O}(N^2)$ 
complexity of \nbody simulations. As we will generally use much less wavefunctions, the complexity is much lower and 
may even trump the usual $\mathcal{O}(N\log N)$ complexity offered by tree-codes or FFT schemes to solve Poisson 
equation. 

An other advantage of working with harmonic wavefunctions to represent the initial conditions is that we have an
intuitive picture of what happens if we remove some modes. In analogy with the Fourier series, the
density will not be represented exactly at every point, but the approximation becomes closer and closer as we include
more and more modes. Knowing some of the properties of the system we want to model may help to get a deeper insight into
which modes are really needed. The same is true when the density is expanded in another basis even if it may be more
difficult to get an intuitive mental picture of the impact of high-order modes when dealing with Chebyshev polynomial
say. 

In many simulations one does not necessarily need the same resolution on all scales. Instead one could work with an
adaptive grid \citep{Plewa2005} and have higher resolution in the scales of interest. This would allow to reach better
precisions while keeping the number of wavefunctions constant. A similar technique is used in \nbody
solvers such as \textsc{RAMSES} \citep{Teyssier2002} or \textsc{ART} \citep{Kravstov1999}.  

In the special case of  simulations of cosmic structure formation, the concept of cosmic variance could help to further
reduce the number of wavefunctions required. Indeed, given that we can only observe one universe, the statistical
fluctuation in large angular patches is high, as not many statistically independent patches are available in our sky.
This is a well-known fact when studying the CMB radiation. This means that the statistical error is
anyway large on these scales, so we do not need to work with a very high precision. Let us also recall that the
freedom of choosing the weight function in the scalar product (\ref{eq:scalar_weight}) of the eigenvalue problem may
help to considerably reduce the number of wavefunctions. Even though this seems to be a promising route to take, we
did not investigate it any further in this work. 

Another area of interest could be the derivation of a scheme to generate initial wavefunctions analytically
in the case of warm dark matter (see for instance \citep{Boyarsky2009}) or for any initial distribution with non-zero
initial velocity spread.

\section{Implementation \& Numerical results}
\label{sec:numerics}

In the previous two sections, we showed how the cosmological Vlasov-Poisson problem (\ref{eq:VP}) can be approximated
by the Schr\"odinger-Klein-Gordon system (\ref{eq:S_KG}). We showed that this approximation is valid in the limit
$\hbar\rightarrow 0$, $c\rightarrow\infty$, $N\rightarrow\infty$. We also demonstrated how the wavefunctions can be
built and that in general they can approximate the true density distribution in the limit $N\rightarrow\infty$. For some
specific cases or for smart choices of eigenfunction basis, the exact $f(x,v)$ can even be ensured with a finite or low
$N$. But let us keep the general case in mind. 

Contrary to the \nbody framework, where the convergence towards the exact solution is not granted in general, we
propose a method where we have a handle on the behavior of the simulation and where we are able to easily test the
dependency of the result on the parameters $\hbar,c$ and $N$. This allows us to truly speak about converged results
and understand the limits of our model. 

Let us now present how this scheme can be discretized and implemented on a computer. We will present the implementation
we used, which is probably the simplest version of what can be done.

\subsection{Implementation}
\label{ssec:implementation}

The simplest possible numerical scheme to solve partial differential equations is to use an explicit scheme in time. An
implicit scheme would be more precise but would require more computing time and memory, the latter quantity being, as we
will show, a rather scarce resource. This explains the choice of an explicit scheme, even if this imposes the use of a
Courant-like condition for our time steps. For the same reasons a scheme accurate up to order $(\Delta \tau)^2$ in time
has been chosen. As going to a precision of order $(\Delta \tau)^4$ would require almost twice as much memory, this
choice can reasonably not be made. Using a symplectic integrator may, however, be useful in future studies as they do
not cost more in terms of memory but conserve the energy of Hamiltonian systems exactly. 

Regarding the spatial derivatives, there are no constraints coming from the memory requirements. One could in
principle go to an arbitrary level of accuracy. But as the time derivatives only have a limited precision, it is not
worth going to a precision higher than $(\Delta x)^4$ , using the usual five-point stencil.

With these two points being set, the system of equations (\ref{eq:S_KG}) can be written on a lattice as follows:

\begin{eqnarray}
 \psi_n(x,\tau + \Delta \tau) &=& \psi_n(x,\tau-\Delta\tau) + i\frac{\hbar\Delta \tau}{a(\tau)} \nabla^2_{\rmn{dis}}
\psi_n(x,\tau) \nonumber\\
&& - i\frac{2\Delta\tau}{\hbar}U(x,\tau)\psi_n(x,\tau) \nonumber \\
U(x,\tau+\Delta\tau) &=& 2U(x,\tau) - U(x,\tau-\Delta\tau) \nonumber\\
&& +c^2\Delta\tau^2\nabla^2_{\rmn{dis}}U(x,\tau) \nonumber\\
&&-\frac{4\pi Gc^2\Delta\tau^2}{a(\tau)}\left(\sum_n^N\lambda_n|\psi_n(x,\tau)|^2-\bar\rho\right),\nonumber
\end{eqnarray}
where the discretized divergence operator is given by

\begin{eqnarray}
 \nabla^2_{\rmn{dis}} f(x) &=& \frac{1}{12 \Delta x^s} \left[\big.-f(x+2\Delta x) +16f(x+\Delta x)\right. \nonumber \\
  && \qquad\left. -30 f(x)+16f(x-\Delta x) -f(x-2\Delta x)\big.\right]. \nonumber
\end{eqnarray}
In the non-cosmological case, the factors $a(\tau)$ can be dropped and one can use time $t$ instead of conformal time
$\tau$. One can show that this scheme is unitary and conserves the norm of each wavefunction. Since the iterative
solution contains the fields at neighbouring lattice sites, care has to be taken that the boundary conditions are
implemented correctly. This is most easily done by augmenting the arrays containing the values of the fields on the
lattice by so-called ghost points to store the periodic boundary conditions.

The last important point regarding the numerics is the choice of $c$ and $\hbar$. It is clear that the Klein-Gordon
equation reduces to the Poisson equation in the limit $c\rightarrow \infty$ and that the higher order terms of 
(\ref{eq:Wigner}) vanish in the limit $\hbar \rightarrow 0$. But numerical stability imposes more conditions on these 
values. An explicit scheme can only converge if there is no information propagating of a distance of one cell during one 
time step. The scalar field propagates at speed of $c$, which gives us the following condition:

\begin{equation}
 \frac{\Delta x}{\Delta \tau} > c,
\end{equation}
which is the usual \emph{Courant condition}.  In practice, the right-hand side
is multiplied by a  constant ($10 - 10^2$) in order to avoid any instability and to remain far from the actual
condition. This condition gives a clear relation between those three quantities and shows that one cannot arbitrarily
improve the spatial discretization without changing the time step size. It is not surprising to have to introduce such
a condition. Indeed, if we were to truly use a value of $c=\infty$ in our simulations, we would have to use smaller
and smaller time steps for a fixed grid spacing. At some point, solving the Poisson equation would become
algorithmically cheaper. The Courant condition is thus the price to pay to avoid solving the usual Poisson 
$\mathcal{O}(M\log M)$ problem.

The evolution of the Schr\"odinger equation also imposes conditions on the time and space slicing. It can be shown that
the following relation 

\begin{equation}
  \frac{\Delta x^2}{\Delta \tau} > \hbar
\end{equation}
must hold, encouraging us, once again, to choose $\hbar$ as small as possible. At this stage, no lower
bound has been analytically derived for $\hbar$. The full dependence on $\hbar$ of the simulation results is still an
open question left for further investigation of this framework.

\subsection{Complexity and memory requirements}
\label{ssec:complexity}

Having presented the algorithm of the time evolution, let us estimate its
computational complexity and memory requirements. Consider a three-dimensional spatial grid made of $M^3$ lattice
points. Let $N_\psi$ be the number of wavefunctions we evolve. Adding the spatial components of the scalar field, $N_f
=
N_\psi + 1$ is the total number of fields we evolve in time. At each time step, we need to compute each of the fields at
every
lattice point, making the algorithm of complexity 

\begin{equation}
 \mathcal{O}(M^3\cdot N_f).
\end{equation}
This has to be compared with \nbody simulations, which have a naive complexity of $\mathcal{O}(N^2)$, that can be
reduced to $\mathcal{O}(N \log N )$ using optimized algorithms. The more particles are tracked, the better becomes the
spatial resolution. Roughly, for a total of $N$ particles, $\Delta x_{\mbox{resol}} \sim L_{\mbox{box}}/N^{1/3}$. In
our case, the spatial resolution is defined by the lattice spacing $\Delta x_{\mbox{resol}} \sim L_{\mbox{box}}/M^3$.
Thus for comparable spatial resolution, we would need $M \sim N^{1/3}$. From this we conclude that the complexity of our
algorithm scales as  $\mathcal{O}(M\cdot N_f)$. In the ideal situation where we only need a few wavefunctions,
$N \sim \mathcal{O}(1)$, our new framework provides an $\mathcal{O}(M)$ algorithm to study structure formation. It
seems that in the worst case we would need as many wavefunctions as there are Fourier modes on the lattice, $N_\psi \sim
\mathcal{O}(M^3)$  implying a complexity $\mathcal{O}(M^2)$, which is the same as the naive force summation in \nbody
simulations. 

These estimates illustrate that our algorithm can indeed compete with the complexity of \nbody simulations. It also
shows how crucial it is to reduce the number of wavefunctions as much as possible. 

Let us next have a look at the memory requirements of our approach. Given that our time evolution relies on a two-level
explicit scheme, we need to keep the field configurations at two time steps in memory. For $N_\psi$ complex
wavefunctions
and one real scalar field components on the whole lattice, we need $2\cdot M^3(2N_\psi+1)$ variables. Assuming that each
is stored as a \texttt{double} of 8 bytes, we can estimate the minimal memory needed by our numerical simulation to be

\begin{equation}
 \geq 2 \cdot M^3(2N_\psi+1)\cdot 8~\mbox{bytes}.
\end{equation}
Let us look once more at the worst case scenario $N_\psi \sim\mathcal{O}(M^3) \sim \mathcal{O}(N_\psi)$. Hence, the 
memory
required now raises to

\begin{equation}
 \geq 32 \cdot N_\psi^2~\mbox{bytes}.
\end{equation}
This has to be compared with \nbody simulations, which have to store at least the position and velocity of each particle
at every time step leading to a memory consumption of
\begin{equation}
 \geq 2\cdot 6 \cdot N \cdot 8~\mbox{bytes}.
\end{equation}
As an example we may give the Millennium simulation \citep{Springel2005}, which needed about 400 GB to store the
information of their $2160^3 \simeq 10^{10}$ particles, in agreement with the above estimate.
We have to conclude that our approach can be strongly constrained by its memory requirements. The gain in computational
complexity seems to have come at a considerable cost in memory. If we consider the 1 TB of
memory available to the Millennium simulation, we could only have $\sim 57^3$ lattice points! However, if we were to
use as many wavefunctions as spatial lattice points, we could as well directly simulate the Vlasov-Poisson system
without introducing any approximation. The whole point of the framework we introduced is to simulate a realistic
probability distribution with a low number of wavefunctions, in which case the memory requirements are not prohibitive
any more and scale with $N$ as in the \nbody case. We also mentioned the idea of using an adaptive mesh to improve the
(spatial) resolution without having to increase the number of wavefunctions.

We now turn to two cases we simulated and show that this new framework is able to reproduce the known solutions. We
also show how the solution depends on the parameters $c$, $\hbar$, $N$ and $\Delta x$.

\subsection{Spherical collapse of a DM sphere}
\label{ssec:collapse}

There are few known non-trivial analytical solutions to the Vlasov-Poisson system (\ref{eq:VP}) even in the static
Universe ($a(\tau) = 1$) case. The collapse of a uniform sphere is among these and is of particular interest for
cosmological applications. A comprehensive treatment of the case, known as \emph{Tolman solution} \citep{Tolman1934},
can, for instance, be found in the textbook \citep{Weinberg1972}. A uniform sphere of initial density $\rho_0$ and 
radius $R_0$ is
collapsing under its own gravitational potential. Gauss's law for gravity states that the evolution of a sphere is not
influenced by the matter lying outside itself. This means that the density inside the sphere will remain constant with
the radius at every time $t$. In other words, all matter will reach the centre at the same time which will lead to
an infinite density. At this stage, the Newtonian description becomes invalid and one would have to use GR in order to
take into account all the effects. In the framework of Newtonian gravity, the matter will simply cross the centre and
oscillates around the centre. Due to the discretization, the simulated central density cannot become infinite and these
oscillations cannot be reproduced exactly. The same shortcomings are present in \nbody codes. 

The evolution of the radius $R$ with time is a quantity which can be easily tracked. In parametric form, the Tolman
solution reads $(0\leq\beta\leq\pi)$:
\begin{eqnarray}
 t &=& \frac{\beta + \sin\beta}{2\sqrt{\frac{8\pi G}{3}\rho_0}}, \\
 R &=& \frac{1}{2}(1+ \cos\beta).
\end{eqnarray}
The density inside the sphere will evolve following the relation

\begin{equation}
 \rho(r,t) = \frac{\rho_0R_0^3}{R^3(t)}.
\label{eq:Tolman}
\end{equation}
For simplicity in what follows, we set $R_0=1$, $G=1$ and $\rho_0 = \pi$. The final collapse time (in arbitrary units)
is then reached when $t_c \approx 0.306$. We will work on the periodic interval $[-5,5]$ which should be big enough to
avoid any unwanted effects from the boundaries.

This problem possesses an obvious spherical symmetry and in order to be able to explore a wide resolution range it is
interesting to re-derive the whole framework presented in the Section \ref{sec:framework} and \ref{sec:IC} using this
assumption. A careful derivation can be found in appendix \ref{sec:spheric} and the end result is that the
Vlasov-Poisson system with spherical symmetry can be re-cast in the one dimensional Schr\"odinger-Klein-Gordon system

\begin{eqnarray*}
 i\hbar \dd{\psi_n}{\tau} &=& \frac{-\hbar^2}{2a(\tau)}\dd{^2\psi_n}{r^2} + \frac{V}{r}\psi_n,\\
- \frac{1}{c^2}\dd{^2V}{\tau^2}+ \dd{^2V}{r^2} &=& 4\pi G r\left(\frac{2\pi}{r^2}\sum_n \lambda_n
|\psi_n|^2 - \frac{4\pi^2\Xi}{V_{\rm{tot}}} \right),
\end{eqnarray*}
where the potential $V = Ur$ and $\Xi$ is the normalization of the wavefunctions (see equation
\ref{eq:normalisation}). The main difference with the framework in presented earlier is the explicit dependency on the
position coordinate $r$. The algorithms developed to find the wavefunctions corresponding to a given distribution
function are identical.

To generate the initial set of wavefunctions and eigenvalues we chose to use the matrix formulation (Section
\ref{ssec:SVD}). The initial density profile being discontinuous, it is obvious that it cannot be recovered exactly with
a finite set of continuous functions. There will be some noticeable differences between the exact density profile and
its approximation appearing at the discontinuity points, that is at the edge of the sphere. It is thus better to use a
approximately correct but continuous density profile. In the case at hand, we used the following initial setup:

\begin{equation}
 \rho(r,t=0) = \frac{\pi}{2}\tanh\left(\xi(r+1)\right) - \frac{\pi}{2}\tanh\left(\xi(r-1)\right),
 \label{eq:smoothed_IC}
\end{equation}
with $\xi=20$. The value of $\xi$ is somewhat arbitrary and has been chosen in order to be as close as possible to
the perfect sphere (i.e. high $\xi$) and avoid any Gibbs oscillation at the edge of the sphere (i.e. low $\xi$). The
results presented here are not really dependent on $\xi$. This parameter has just been introduced for convenience
and to avoid having to analyse the effects of these unwanted and unrealistic oscillations. In fact, even a value of
$\xi=\infty$ yields comparable results to what is shown below once the Gibbs oscillations have been smoothed out
manually from the output.

\begin{figure}
\includegraphics[width=84mm]{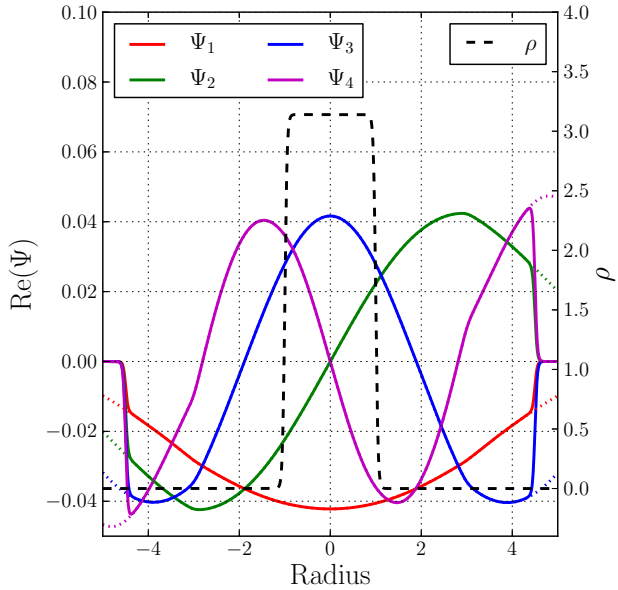}
 \caption{The first four wavefunctions contributing to the WDF of the approximate uniform sphere before
(dotted lines) and after (superimposed solid lines) having applied the smooth window function to make them vanish at the
boundaries of the box. These functions are different from zero almost everywhere but their combination in a WDF
corresponds to the density profile (dashed black line, equation \ref{eq:smoothed_IC}), which is zero on most of the
interval.} 
 \label{fig:wavefunctions}
\end{figure}

Once discretized on a lattice, the eigenvalue decomposition is straightforward to obtain, for instance using the SVD
function implemented in the usual scientific software packages. Recall that there is no guarantee
that the obtained functions will be periodic on the interval of interest or even
that these function will be smooth. It is a pure matrix operation without any
relation between the matrix elements representing the wavefunctions. The interval $[−5, 5]$ has been uniformly
discretized regularly in $5000$ line elements in order to get a high enough spatial accuracy. This means that we 
want to perform the SVD decomposition of a $5000 \times 5000$ matrix and that we can use up to $N = 5000$ wavefunctions 
in the simulation. The matrix reads

\begin{equation}
 \hat G_{ij} = \rho\left(\frac{r_i+r_j}{2}\right),
\end{equation}
where the $r_i$'s are the uniformly distributed lattice points. This matrix is by construction symmetric and positive
definite, meaning that its eigenvalues will be positive or null. Most of the SVD routines in scientific packages sort
the eigenvalues $\lambda_n$ according to their magnitude which allows us to classify the most important contributions 
and discard the negligible terms in equation \ref{eq:WDF} if one does not want to use all the $N$ functions. The first 
four wavefunctions are shown on  figure \ref{fig:wavefunctions}. 

The wavefunctions obtained through this procedure are smooth (at the lattice level at least) and real but are not
periodic nor anti-periodic, which leads to spurious diffusion at the boundaries of the box. For this reason, we decided
to multiply them by a square-box like compact function going to zero close at the box boundaries. The first four
wavefunctions before and after applying this window filter are also shown on  figure \ref{fig:wavefunctions}. This 
procedure does not modify the distribution function obtained through the WDF. This reflects the fact that there is 
infinitely many ways to decompose the same $f(r,v)$ in wavefunctions. Notice that this procedure of adding a window 
function can only be done if the density vanishes at the boundaries. 

\begin{figure}
\includegraphics[width=84mm]{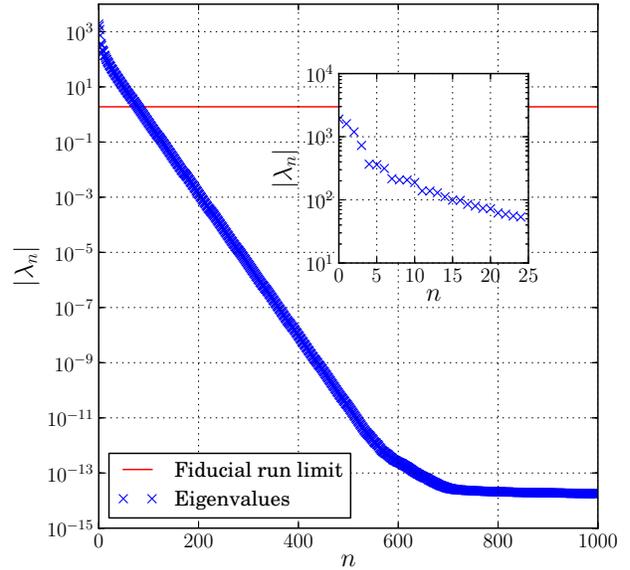}
 \caption{The first 1000 eigenvalues $\lambda_n$ corresponding to the SVD decomposition of the spherical collapse
problem. The values decrease rapidly and become negligible (when compared to the first one) for $n > 100$. They even
reach a minimum close to the machine epsilon for $n>700$. Our fiducial run uses all the wavefunctions up to $n=79$
which
corresponds to $\frac{\lambda_n}{\lambda_0} > 10^{-3}$. This limit is shown as the red solid line on the figure. The
small panel
presents a zoomed-in region of the eigenvalues with $n<25$. The decrease on this small subset is already of more than
an order of magnitude.} 
 \label{fig:eigenvalues}
\end{figure}

Apart from the wavefunction, the eigenvalue associated to each mode also enters the WDF (equation \ref{eq:WDF}). These
are obtained at the time than the discretized wavefunctions and their values are represented on figure
\ref{fig:eigenvalues}. The actual normalization of the eigenvalues does not really matter as any common factor can be 
absorbed
as normalization in front of the WDF. But the ratio of the values plays a role. All the different wavefunctions
(modes) entering the decomposition of $f(r,v)$ may not play an important role exactly as in the case of a Fourier
series decomposition where some of the modes can safely be neglected. As can be seen on figure \ref{fig:eigenvalues},
the values of the various $\lambda_n$ decrease rapidly and for $n>100$, they represent less than $10^{-3}$ of the
most important mode. As the eigenvalues are constants of motions, we can hope that neglecting modes with a high $n$
(and hence a small $\lambda_n$) will not affect the simulation too much. In fact, unless the magnitude of
the wavefunction corresponding to one of the neglected mode grows significantly over the course of the simulation,
this mode should remain small at all time and can thus be safely ignored.

In our main run, we used all eigenfunctions $\Psi_n$ whose eigenvalue fulfils $\lambda_n>10^{-3} \lambda_0$,
which left us with only $N=79$ functions to evolve. The other numerical parameters we chose in our fiducial
run are $c=10$ and $\hbar=0.005$. We do not expect $\hbar$ to have a big impact on the results in this case as the
potential is a combination of a second and third order polynomial for which the higher order corrections in the Wigner 
equation (\ref{eq:Wigner}) should be small.

\begin{figure*}
\includegraphics[width=\textwidth]{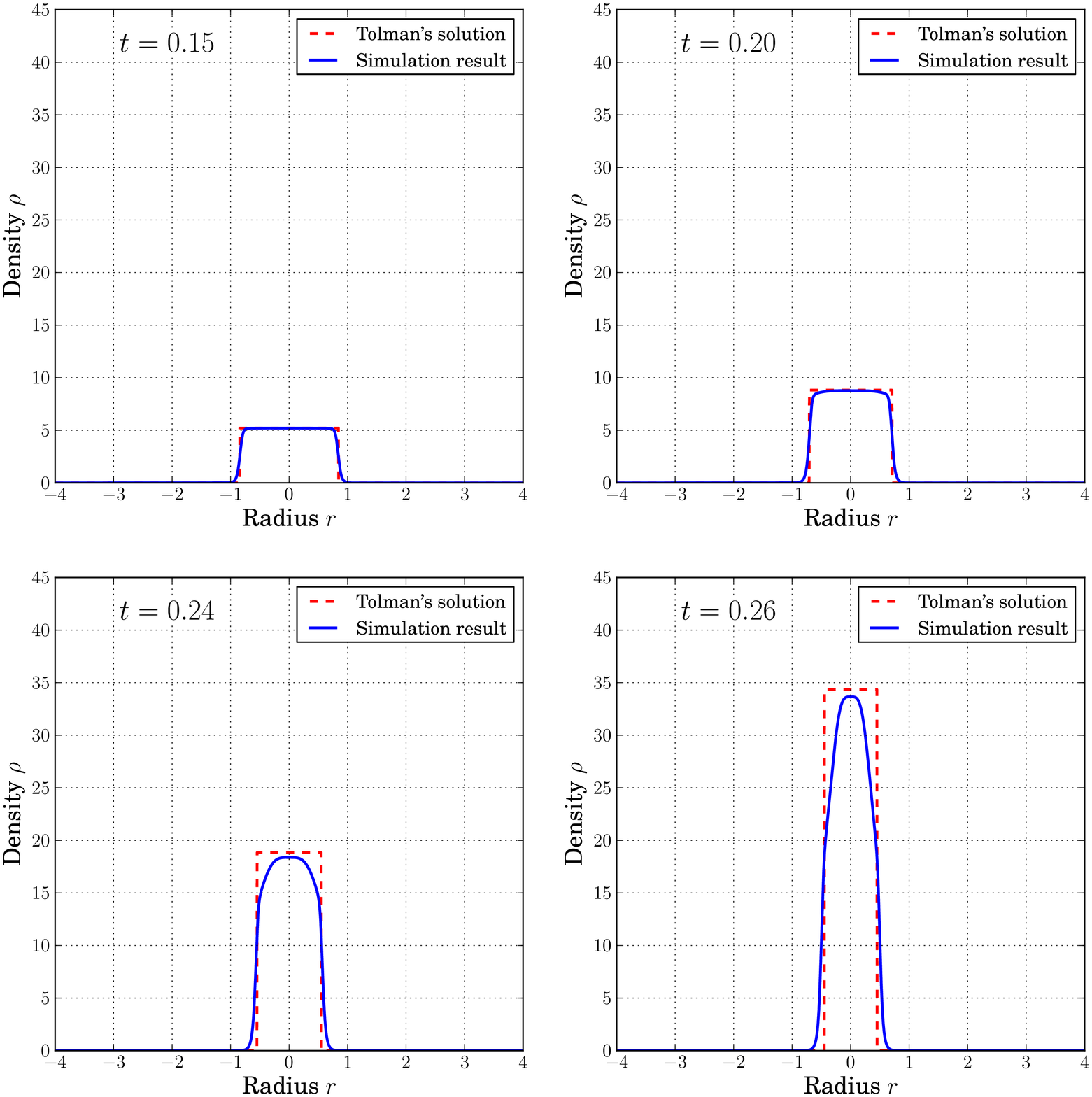}
 \caption{Four different density profiles (blue solid line) of our fiducial run at $t=0.15$, $t=0.20$, $t=0.24$ and
$t=0.26$
together with the analytical solution (red dashed line) at the same time steps. The simulation follows
almost perfectly the analytical solution until $t\approx 0.20$ and then starts to deviate. For obvious reasons, the
situation gets worse at later times and even the centre of the sphere does not follow the exact solution any more. After
the collapse time $t_c \approx 0.306$ the behaviour becomes clearly non-physical due to the unresolved infinities. The
very
centre of the sphere still follows Tolman's solution closely on the last two panels but the sharp features at the edge
of the sphere get increasingly more difficult to represent. This suggests that increasing the spatial resolution of the
lattice may help getting better derivative estimate and hence improve the quality of the result.} 
 \label{fig:rho_evolution}
\end{figure*}

Figure \ref{fig:rho_evolution} shows four density profiles at different time steps in the simulation together with the
analytical solution (equation \ref{eq:Tolman}). Until $t\approx 0.2$, the behaviour of the density profiles remains
close to the exact solution apart from the very edges of the sphere that are slightly smoothed. The centre of the 
density profile is almost flat as expected and has almost the correct value. When coming closer to the collapse time
$t_c\approx0.306$, the profiles starts to deviate more and more from the expected profile. This can be seen on the last
two panels of figure \ref{fig:rho_evolution} where the density inside the sphere is clearly different from a square box
function. The very centre of the sphere still remains close to the analytical solution but the edges are not sharp
any more and are smoothed over many lattice elements. This strongly suggests that the estimation of the derivatives of
both the potential and the wavefunctions are getting poor or that the number of wavefunctions used in the run is not
high enough. Our scheme uses a fourth order accurate derivative stencil but this does not necessarily help recovering
sharp features such as the one present at the edge of the sphere. Increasing $N$ and reducing $\Delta r$ may help
recover the right density profile everywhere in the sphere. 

The results on figure \ref{fig:rho_evolution} have been obtained using $N=79$ wavefunctions corresponding to all
eigenvalues $\lambda_n > 10^{-3}\lambda_0$. This should be sufficient as the eigenvalues are constants of motion and we
do not expect any of the neglected wavefunctions to grow by a huge factor over the course of the simulation. In order
to assess this, we run the same simulation with $N=155$, corresponding to all wavefunctions whose eigenvalues
$\lambda_n > 10^{-5} \lambda_0$. Notice here that decreasing the minimal eigenvalue entering the WDF by two orders of
magnitude only increases $N$ by a factor of $2$. We are thus far from the worst case scenario (see Section
\ref{ssec:complexity}) where the same number ($N=5000$) of wavefunctions than lattice points have to be used. 

\begin{figure}
\includegraphics[width=84mm]{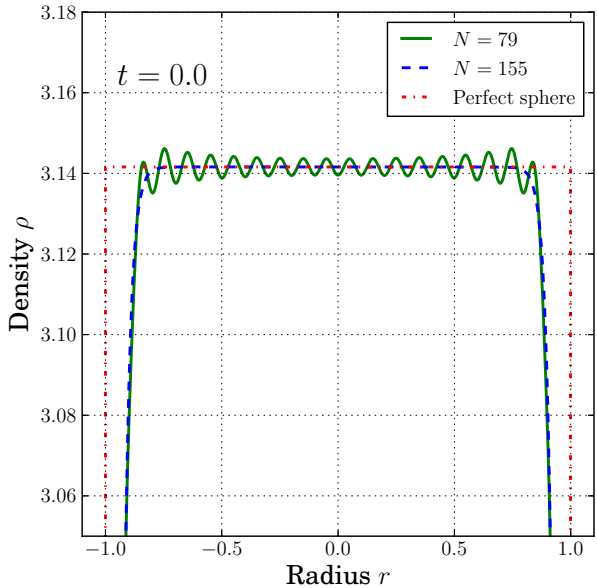}
 \caption{Comparison of the density profiles at the initial time for $N=79$ (green solid line) and $N=155$ (blue dashed
line) wavefunctions. The figure zooms in the central regions where the difference can be spotted. The $N=79$ line
presents a lot of oscillations that are suppressed if more wavefunctions are used. The $N=155$ line almost perfectly
matches the density profile given by equation \ref{eq:smoothed_IC}. Notice, however, that this differs from the
perfect sphere profile (red dashed-dotted line), which cannot be represented by a finite set of continuous functions.}
 \label{fig:comparison_N}
\end{figure}

Figure \ref{fig:comparison_N} shows a comparison at $t=0$ of those two initial setups. The figure only shows a
zoomed-in view focused on the sphere itself as the difference are less visible in the outer regions of the
simulation domain. As can be seen, the $N=155$ initial setup (dashed blue line) is a much better representation of the
smoothed density profile (equation \ref{eq:smoothed_IC}). At this resolution, the two are indistinguishable. The $N=79$
initial conditions (green solid line) presents some oscillations inside the sphere that are very similar to the Gibbs
phenomenon that appears when computing the Fourier series of the square box function. Using a smoothed density profile
and decomposing in eigenvalues using the matrix formulation thus yields a result which is very similar to generating
the ICs through the Fourier decomposition (Section \ref{ssec:fourier}). This could have been anticipated by looking at
the wavefunctions (figure \ref{fig:wavefunctions}), where the different $\psi$'s resemble sines and cosines functions
at least qualitatively. As can be seen, the relative error introduced by using only $N=79$ wavefunctions is of the
order $10^{-3}$, whereas the error computed when using $N=155$ is smaller than $10^{-6}$, showing once again that 
increasing the number of eigenfunctions used by a factor of $2$ increases the simulation by more than $2$ orders of 
magnitude. However, it should be noticed that using another basis or weighting function for the eigenvalue decomposition
(\ref{eq:scalar_weight}) may yield another $N$ with the same or different accuracy. Comparing the number of
wavefunctions only makes sense when using a similar decomposition technique. Let us also mention that we also tried
using harmonic functions and Chebyshev polynomials for this test case and obtained similar results.

At later times, the simulation snapshots are identical to the ones presented earlier on figure \ref{fig:rho_evolution}.
The relative difference between the two runs is of order $10^{-3}$ as in the initial conditions. This implies that the
difference between our simulation results and the analytical solution can not be reduced by using more and more
wavefunctions. The additional modes that have been discarded when using only $79$ eigenfunctions do not contribute
significantly to the final results. This could have been expected as their weightings ($\lambda_n$) are very small
compared to the main modes. We can thus gain confidence in the way we generate ICs, discarding higher order modes may
not be an issue and we may be able to run our algorithm in a near linear regime even when a violent collapse of matter
is studied. 

In conclusion, increasing $N$ does make the initial conditions and the simulation outputs converge towards a solution at
a high rate. However, the discrepancy between the solution and the simulation does apparently not come from the wrong
choice of the parameter $N$. Let us now explore the dependency on the grid resolution. 

\begin{figure}
\includegraphics[width=84mm]{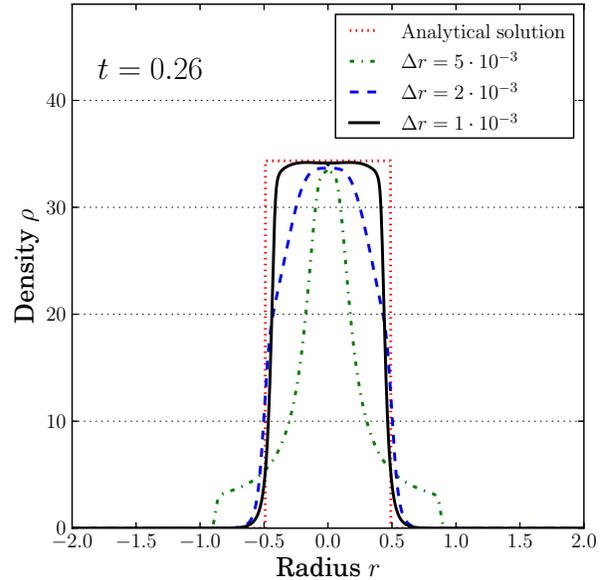}
 \caption{The output at $t=0.26$ for different lattice resolution using $N=79$ wavefunctions. The green dash-dotted
line corresponds to a low resolution run with $\Delta r=5\cdot10^{-3}$, the blue dashed line corresponds to our
fiducial run at $\Delta r = 2\cdot 10^{-3}$ and the black solid line is the output of a high resolution run using
$\Delta r=10^{-3}$. The quality of the output is clearly improved by using a higher resolution lattice. This can be
directly related to the problem of estimating sharp derivatives on a grid, where the only solution is to increase the
resolution.} 
 \label{fig:comparison_dr}
\end{figure}

On figure \ref{fig:comparison_dr}, we show the results of three runs at different grid resolutions leaving the number
of wavefunctions and all the other parameters fixed. The blue dashed line corresponds to the fiducial run ($\Delta
r=2\cdot10^{-3}$), the green dotted line to a lower resolution run using $\Delta r=5\cdot10^{-3}$ and the black solid
line corresponds to the high resolution run with $\Delta r=10^{-3}$. As can be seen, increasing the resolution has a
huge impact on the quality of the result. As anticipated, the sharp features can only be resolved correctly when enough
grid points are used. Notice that the high resolution run almost matches exactly a rescaled version of the initial
density profile (equation \ref{eq:smoothed_IC}), but does break down at later times in the same way that the fiducial
run did between $t=0.20$ and $t=0.26$ (figure \ref{fig:rho_evolution}). Increasing the resolution is thus important to
be able to retrieve all features of this somewhat artificial test case. This test case presents a strong density 
gradient at the edge of the sphere which does not spread over many cells. This makes it difficult to resolve for a grid 
code but in a cosmological simulation such sharp gradients should not arise as the density profiles usually follow power 
laws and do not have infinite gradients. As already mentioned, using an adaptive mesh would help in such a case as more 
resolution elements could be used at the edge of the sphere without having to slow down the simulation due to an 
unnecessary oversampling of the steady regions.

This demonstrates that our framework converges towards the analytical solution once the spatial resolution is high
enough and once the number of wavefunctions has been carefully chosen to represent the distribution function 
of interest.

\begin{figure}
\includegraphics[width=84mm]{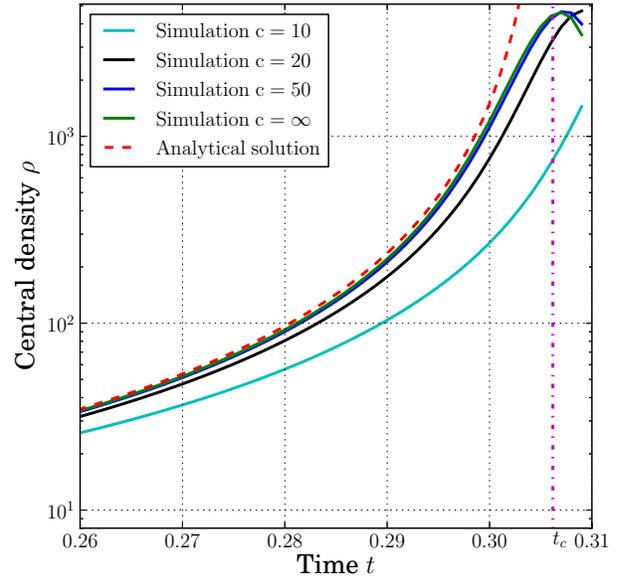}
 \caption{Evolution of the density at the centre of the sphere ($r=0$) for different values of the numerical speed of
light $c$. The red dashed line corresponds to the analytical solution (\ref{eq:Tolman}), the vertical dash-dotted line
represents the final collapse time and the different solid lines correspond to the different values of $c$. The higher
the value of $c$, the closer the line lies to the exact solution. The line with $c=\infty$ has been obtained by
solving Poisson's equation on the grid instead of evolving gravity using Klein-Gordon's equation. The quality of the
simulation outcome clearly depends on the value of $c$ but if the value is high enough (compared to the velocity of
the matter), the difference with the $c=\infty$ becomes very small. 
Once the density peak has been reached, the value of $\rho(r=0)$ decreases as is expected after the different matter
shells have crossed. The simulations have not been carried on much beyond this point as the departure from the
analytical solution is already significant. Moreover, the peak can not be represented accurately by any numerical mean
and any subsequent event would be erroneous.} 
 \label{fig:comparison_c}
\end{figure}

This new framework should converge towards the solution in the limit $N\rightarrow\infty$, $\Delta x\rightarrow 0$,
$c\rightarrow\infty$ and $\hbar\rightarrow0$, the last two being, despite their physical origin, only numerical
parameters. Figure \ref{fig:comparison_c} presents the evolution of the density at the centre of the sphere for our
fiducial run and for higher values of $c$. The simulation with $c=\infty$ has been obtained by solving Poisson's
equation on the grid at every time step instead of using Klein-Gordon's equation. Increasing $c$ improves the quality
of the result and even relatively small values ($c=50$) of this parameter lead to a behaviour close to the limiting
case. Poisson's equation can thus safely be replaced by Klein-Gordon's equation.The maximal speed reached by matter 
shells in our fiducial run is $p\approx10$ before the very end of the collapse, which can anyway not be studied by a 
simulation. Using a value of $c=10$ is thus intuitively too low and this plot confirms this. The speed of gravity must 
be at least a few times bigger than the matter velocity.

Once the peak has been reached, the different matter shells should cross the centre and the density at $r=0$ has to
decrease. The start of this behaviour can also be seen in figure \ref{fig:comparison_c}. The main issue with this
analysis is that is happening after the moment where the density at the centre becomes infinite and hence not
representable on a computer. In practice, all the wavefunctions should become infinite at this precise point and zero
elsewhere. This is obviously impossible on a lattice and does anyway lead to inaccurate derivatives. To get closer and
closer to the singularity requires a finer and finer mesh. The smaller the mesh size, the better the shell crossing can
be followed. 

Notice, however, that this is an issue present in this ideal sphere case only. In a realistic scenario, where the
matter has a non-zero radial velocity and in an expanding background, the usual NFW profiles \citep{Navarro1996} should
be recovered without singularity problems. This would, however, require a truly 3D simulation and not just a spherically
symmetric 1D setup.

Increasing $c$ has a big impact on the simulation run time as the time step size varies as $c^{-1}$, making the total
simulation wall clock time proportional to $c$. An option that has not been explored here is to change the value of $c$
to be always a (small) multiple of the maximal matter speed in the simulation. This would allow us to choose bigger
time steps in the early stage of the simulation when all the matter moves slowly. It would also avoid making an initial
guess for the value of $c$ without knowing how fast the matter will move during the run. 

As discussed earlier, the dependency on $\hbar$ is difficult to test in this case as the analytical potential only
presents first order corrections in the Wigner equation. We did run some simulations 
with various values of this parameter without noticing important differences in the behaviour of the matter 
distribution. Understanding the exact dependency on $\hbar$ of the framework is left to a future work.

This simple spherical collapse test showed that we were able to reproduce the analytical solution in the limit
$N\rightarrow\infty$, $\Delta x\rightarrow0$ and $c\rightarrow\infty$ as expected. We investigated the different
deviations from the exact solution and could explain them through our choices of numerical parameters. We also
discussed how the implementation could be improved by using a mesh-refinement and adaptive $c$ values. The results
obtained so far show that this new framework can reproduce known solutions and give us confidence to use it on more
involved cases.

\subsection{Going beyond the first collapse}
\label{ssec:nonLinear}

With the previous test case, we showed how our framework was able to reproduce the collapse of a matter distribution in 
the linear regime and studied the dependency on the model parameters. However, in most cases of interest, the systems 
considered in simulations are way past the linear regime. They also present multiple matter streams, i.e. at a given 
position $x$, there are multiple velocities $v$ and the distribution function is ``wound up''. It is hence important to 
explore whether this behaviour can be recovered by our framework. Note that tracking precisely these multiple matter 
streams is extremely difficult in the case of \nbody simulations unless advanced phase-space tessellation techniques 
are used \citep{Abell2012,Shandarin2012}.

The test case presented in the previous section exhibits a nice analytical solution but, as discussed, the matter 
distribution becomes infinitely thin at the time of the collapse which makes all attempts at taking derivatives 
difficult. To alleviate this issue, we use a simpler one dimensional test case with a much smoother density 
distribution. In this section, we study the evolution in one dimension of the cold distribution function

\begin{equation}
 f(x,v) = \rho(x)\delta(v), \qquad \rho(x) = \rho_0\exp(-x^2 / 2s^2),
\end{equation}

with $s$ the scale size of the matter distribution. This test case has already been studied by \citep{Widrow1993} in 
the context of their framework which makes use of a Husimi distribution instead of the Wigner one. We will use a 
periodic domain of size $L_{\rm{box}} \gg s$. 

The first step in the algorithm is to decompose the initial condition into a series of wavefunctions. There are many 
ways to do this and one could easily use either a decomposition in terms of sine waves or using the matrix 
decomposition used in the previous test case. The decomposition in Fourier modes is straightforward and the initial 
distribution function can be recovered in a satisfactory way with less than $10$ wavefunctions. However, to demonstrate 
the fact that the number $N$ of wavefunctions is only a relevant quantity once a decomposition scheme has been chosen, 
we will use a simpler single wavefunction to represent $f(x,v)$:

\begin{equation}
 \Psi(x,t=0) = \sqrt{\rho(x)}.\label{eq:3}
\end{equation}

Using this simple decomposition leads to a an initial Wigner distribution of the form

\begin{equation}
 f(x,v) = \rho(x)\exp(-v^2/2\hbar^2),
\end{equation}

once equation \ref{eq:initial_distribution} has been applied. This example also explicitly shows how $\hbar$ enters the 
framework and the effect this quantity has on the initial conditions and hence on the subsequent evolution of the 
distribution function. The previous test case gave us some insights into how to choose the value of the speed of light 
$c$. We could use similar considerations here to choose an appropriate value, however, to simplify the discussion, we 
choose to set $c=\infty$ and solve Poisson's equation for gravity at every time step using the fast Fourier transform 
algorithm. In what follows, we set $\rho_0=1$, $s=10^{-2}$, $\hbar=10^{-3}$, $L_{\rm{box}}=1$ and discretize our volume 
in $M=100$ 
intervals. 

\begin{figure}
\includegraphics[width=84mm]{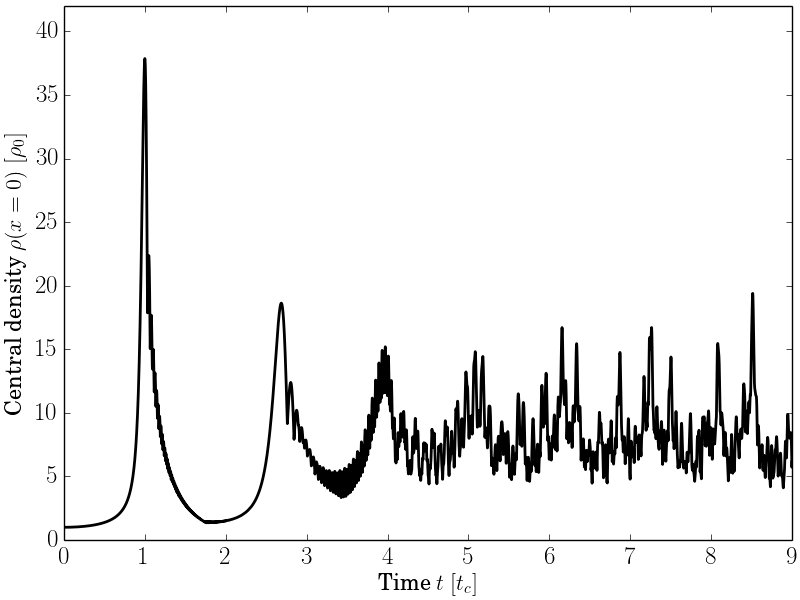}
 \caption{Evolution of the density at the centre of the domain ($x=0$) beyond the first collapse ($t>t_c$). The 
multiple collapses 
and re-expansions of the matter distribution can be tracked by the framework. The appearance of multiple matter streams 
during the evolution of the collapse can be resolved by the simulation even with one single wavefunction.}
 \label{fig:nonLinearEvolution}
\end{figure}

To trace the non-linear evolution of the system, we trace the value of the density field at $x=0$. This is in essence 
similar to figure \ref{fig:comparison_c} for the previous case but we now let the simulation run past the initial 
collapse time. The result of this evolution is shown on figure \ref{fig:nonLinearEvolution}. As can be seen, the first 
peak is followed by a relaxation of the system and then by a series of additional regularly spaced collapses that occur 
every time the matter distribution crosses the spatial origin. The first four peaks can be well followed despite the 
relatively low spatial resolution and the single wavefunction used in this example. Using a higher value of $M$ leads 
to more peak being resolved and less oscillations in the value of the central density. This, once again, highlights the 
key importance of the spatial resolution over the raw number of wavefunctions. This is also true in standard \nbody 
simulations that use meshes to solve Poisson's equation. The quality of the solution is mostly driven by the high 
number of grid elements and less by the pure number of particles used in the simulation. Our framework is hence able to 
track the collapse of a matter distribution when multiple shell crossings occur and in the presence of multiple 
matter streams.\\

It is interesting to discuss what would happen if more wavefunctions were used to represent $f(x,v)$. Obviously, one 
cannot add more $\Psi$ to the decomposition given by equation \ref{eq:3} as it already provides a exact match to the 
density profile; any addition would reduce that agreement. Note that one might want to consider doing so as it could 
reduce the spread in velocity and hence give a better set of initial conditions but there does not seem to be a simple 
way to do so. Alternatively, one might consider using a Fouried decomposition of $\rho(x)$ (section \ref{ssec:fourier}) 
with a high enough number of cosine waves to reproduce $\rho(x)$. A small number of waves will be sufficient as the 
case is smooth enough and by doing so, the spread in momentum can be reduced. The more wavefunctions are used, the 
smaller the initial spread in velocity, allowing us to get rid of the explicit dependence on $\hbar$ in the initial 
Wigner distribution. This obviously comes at a higher numerical cost but might be necessary in some situations. The 
freedom of getting a spread in velocity space smaller than $\hbar$ is a fundamental difference between or framework and 
earlier work based on the Husimi function \citep{Widrow1993, Davies1997}.

\subsection{Linear structure growth in \lcdm}
\label{ssec:growth}

We now apply this new framework to a simple example of cosmic perturbation growth. We will consider the simplest
possible case of a constant background $\bar\rho$ in a cold dark matter Universe and a small perturbation $\epsilon \ll
1$ with a single Fourier mode $k_p$ taken along the $x$ direction:

\begin{equation}
 \rho(\vec{x},t) = \bar\rho + \epsilon\left[\cos(k_p x_x) + \sin(k_p x_x) \right].
\end{equation}
This basic setup should be sufficient to study the behaviour of the framework in an expanding Universe case. 

Generating the wavefunctions corresponding to this initial distribution function was discussed in Section
\ref{ssec:fourier}. The equations (\ref{eq:Fourier_decomposition}) define a representation of the density in
terms of wavefunctions. As we only have one single mode, we only need one wavefunction for the constant background
($\psi_0$) and two for the perturbation. We run the simulation on a $30^3$ spatial lattice corresponding to a physical
box size of $60~\rm{Mpc}$. It is important to notice here the low number of wavefunctions $N = 3 \ll 30^3$, allowing us
to run our algorithm in a near linear regime. We choose the scale of the perturbation to be larger than the Nyquist
frequency and small compared to the box size to avoid unwanted effects due to the limited box size. 

In a purely matter dominated (Einstein-de Sitter) Universe, the scale factor $a(\tau)$ will grow as the square of the
conformal time. Without loss of generality, we can normalise it such that it is equal to one at the start of the
simulation $a(\tau_{\rm{ini}})=1$, implying

\begin{equation}
 H^2_{\rm{ini}} = \frac{8\pi G}{3} \bar\rho_{\rm{com}} a(\tau_{\rm{ini}})^{-3} = \frac{8\pi G}{3} |\psi_0|^2.
\end{equation}
 The above relation fixes the value of this wavefunction in terms of the initial Hubble parameter, which can be
computed
by rescaling today's value $H_0$ to the redshift corresponding to the beginning or our simulation

\begin{equation}
 H_{\rm{ini}} = H_0(1+z_{\rm{ini}})^{3/2}.
\end{equation}
We ran our simulations for the choice $z_{\rm{ini}} = 1000$ and using today's Hubble parameter $H_0 \simeq
70~\rm{km}~\rm{s}^{-1}~\rm{Mpc}^{-1}$. The initial conditions with a density contrast of $\delta_{\rm{ini}} =  10^{-6}$
where evolved up to a redshift of $z_{\rm{fin}} = 200$. We use a normalised time line such that
$z_{\rm{ini}}$ corresponds to $ \tau=0$ and $z_{\rm{fin}}$ corresponds to $\tau=1$ using  $3\cdot 10^4$ time steps. The
same initial perturbations were evolved in a matter-dominated, expanding universe and in a static universe without
expansion. 

The parameter $c$ has been chosen in accordance with the results of the previous test by making it bigger than the
speed of the matter in the simulation and small enough to avoid drastically pulling down the time step. In what
follows, $c=10$. The parameter $\hbar$ has been, once again, chosen small enough for the quantum corrections to be
negligible. More specifically, this means that the first quantum correction in the Wigner equation (\ref{eq:Wigner})
has to be small compared to the contribution to the classical Vlasov equation:

\begin{equation}
 \dd{V}{x}\dd{P_W}{v} > \frac{1}{24}\hbar^2\dd{^3V}{x^3}\dd{^3P_W}{v^3}.
\end{equation}
We verified that this indeed the case in our simulations when using $\hbar=0.005$. We could, in principle, also verify
that the higher-order corrections are also suppressed but computing the fifth derivative of the potential will lead to a
very noisy estimate and may not lead to useful results. 

\begin{figure}
 \includegraphics[width=84mm]{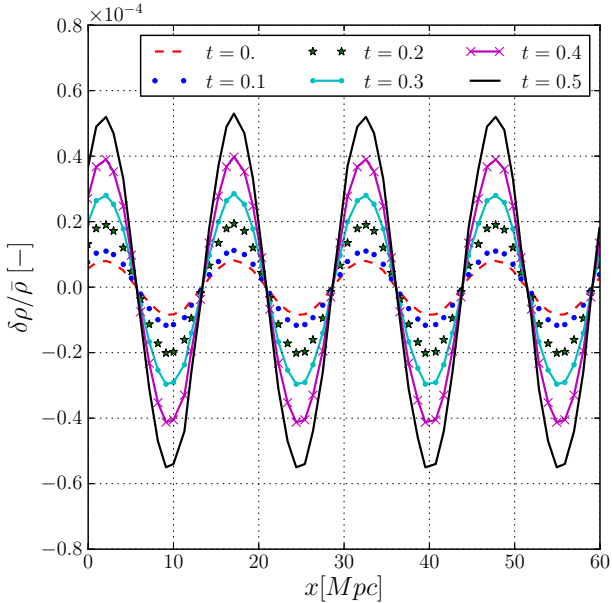}
 \caption{Time evolution of the density field in an expanding universe. The different lines correspond to various
time steps in normalised units. The values are taken along one line parallel to the $x$-axis in the box but all lines
yield the same results. The initial amplitude of the harmonic density increases with time, without distortion of the
shape, as expected from structure formation in the linear regime.}
 \label{fig:density_growth}
\end{figure}

Figure \ref{fig:density_growth} shows the time evolution of the density. The initial amplitude of the harmonic density
increases with time, without distortion of the shape, as expected from the linear regime of structure formation. The
growth of structure seems thus to be well reproduced by our framework even with such a low number of lattice points and
wavefunctions. The simulation could, in principle, be carried on to a much lower redshift than $z=200$ but at some
point, the spatial resolution issues highlighted in the previous test would appear here as well. Recall that we have
only 30 grid points in our $60~\rm{Mpc}$ box. As soon as the variation of the density becomes important on a scale of
order a few $\rm{Mpc}$, the discretized derivatives will cease to approximate the analytical ones and our formalism 
will break down as would any uniform grid code with the same resolution. We, hence, decided to restrict ourselves to the 
regime where our density field and the wavefunctions are well behaved in order to make a useful analysis of the 
results.

\begin{figure}
 \includegraphics[width=84mm]{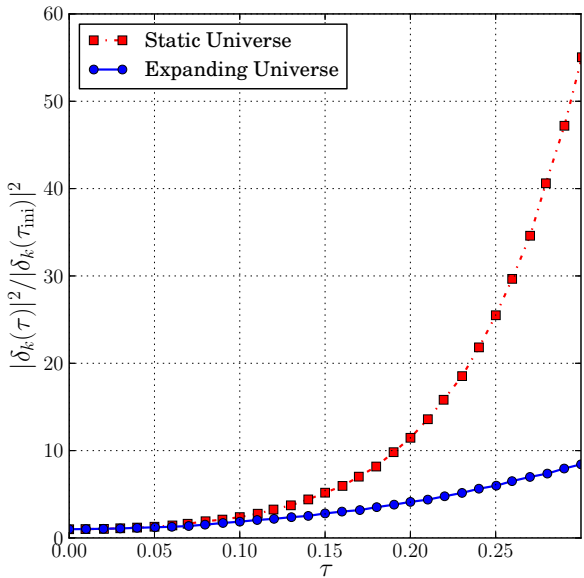}
 \caption{Comparison of the growth of the perturbation $|\delta_k(\tau)|^2/|\delta_k(\tau_{\rm{ini}})|^2$
in a non-expanding universe (red squares, upper line) and in a matter-dominated, expanding universe (blue circles, lower
line) as a function of the conformal time $\tau$. As expected, the growth is clearly suppressed in the presence of
expansion.}
 \label{fig:growth_comparison}
\end{figure}

To analyse the growth of the perturbation in more detail, we performed a Fourier transform on the density contrast to
obtain $|\delta_k|^2$. In this way we could also check that no other Fourier modes than the one initially present were
excited during the simulation. This is a cross-check for the linearity of the evolution of the small density
perturbation. The figure \ref{fig:growth_comparison} compares the growth
$|\delta_k(\tau)|^2/|\delta_k(\tau_{\rm{ini}})|^2$ for our mode in the expanding and non-expanding universes. Clearly,
the growth of the perturbation is suppressed in presence of expansion.

These results clearly show that our framework is able to follow the growth of a single-mode density perturbation in an
expanding background. The main features are recovered even when a low number of lattice points and wavefunctions is
used. By taking advantage of the ease of decomposition in orthonormal Fourier modes of the cosmological power spectrum
(Section \ref{ssec:cosmo_IC}) more complex cases can be studied by superposing the different modes. The results
obtained here give us confidence about the behaviour of the framework in the non-linear regime of cosmic growth. The
main features of \lcdm can probably be recovered in a higher-resolution run with more wave functions and a
longer run time.

As in the previous test case, one could track the matter distribution into the non-linear regime and track the 
appearance of multiple matter streams. This is of course of crucial importance for realistic simulations of structure 
growth in the Universe. It is, however, obvious that the addition of the scale factor $a(\tau)$ in the simulation will 
not alter the behaviour seen in section \ref{ssec:nonLinear} and we are confident that multiple streams would also 
appear and be correctly tracked by the evolution of the wavefunctions. A more detailed study of the framework in the 
context of cold or warm dark matter cosmologies is left for future work.

\section{Conclusion}
\label{sec:conclusion}

We introduced a new alternative framework for simulation of structure
formation which is not based on the usual discretization of the density field
in a set of particles. We made use of the Wigner distribution function to
recast the distribution function in a set of wavefunctions. We could thus
replace the 6-dimensional Vlasov equation by a set of Schr\"odinger equations
acting on the wavefunctions. The Poisson equation for gravity has been
transformed into a Klein-Gordon equation making the system of equations
completely local. We demonstrated how this system of equation could be derived
from a Lagrangian and how the total energy and mass are conserved by the
equations of
motion.

We presented different methods to generate the initial conditions depending on
the distribution function of interest and described how a cosmological power
spectrum can be discretised in a low number of wavefunctions. The framework
has then be tested on two simple models to assess its validity and the
dependency of the outcome on the numerical parameters has been sketched. The
results obtained thus far show that this framework is viable and may become a
possible alternative
to the \nbody method. 

The important new features introduced in this framework are the possibility to
simulate a generic distribution function and not only cold dark
matter. Although finding an easy and generic way to generate initial
conditions for warm or hot dark matter remains an open question, there are no
intrinsic limitations in the framework that could prevent such simulations. It
also provides an alternative to \nbody codes and could thus help assess
the validity of simulations. Our technique can be shown to converge towards
the solution in the limit $c\rightarrow\infty$, $\hbar\rightarrow0$ and
$N\rightarrow\infty$ making the formal convergence studies possible. The
computational complexity of the algorithm grows as $\mathcal{O}(N\cdot M)$
where $M$ is the number of lattice points. This demonstrates the importance
of finding the appropriate decomposition of the distribution function in
wavefunctions. The complexity can hence be anything between linear and
quadratic in the number of points. The case of structure formation
may be close to the ideal case thanks to the possibility to discretise the
power-spectrum in a low number of modes. 

This scheme is especially aimed at tackling the fundamental challenges that the \nbody method faces when dealing with non-CDM cosmologies. 
This includes simulation of a WDM Universe but also neutrino components in a standard \lcdm model or any other particle with non-negligible thermal velocities. 
At the same time, exploring CDM through this framework might help understand more precisely the limitations of the \nbody method by comparing results in the same way that various
hydrodynamic solvers help understand the behaviour of the codes and their limits.

One could also argue \citep{Sikivie2010} that such an approach may be
appropriate to simulate axions which remain quantum during the entire
cosmological evolution. In such a case, the real value of $\hbar$ and particle
mass would have to be used, which would, however, probably lead to very high
computational costs.

In this paper, we presented the validity of the method but many promising and interesting options have not yet been
explored. The first obvious domain to investigate is the dependency on $\hbar$ of the results. Early results tend to
show that it may not be a crucial issue thanks to the universal gravitational profiles being low-degree power laws and
hence generating only small quantum corrections to the Vlasov equation. It still remains an open question. 

The other important area of investigation is the generation of initial conditions for more general cases than simple
CDM. The procedures presented here can not be applied without making some educated guess on the best shape of harmonic
functions or without having to solve gigantic matrix eigenvalue problems. Combining some of these procedures or using
interpolation techniques between lattice points are possible improvements worth exploring.

Finally, on the implementation side, lot of work can be done to make the codes more efficient. We already discussed the
possibility of using an adaptive mesh to refine the grid in the regions of interest. It may also be possible to use an
adaptive value of $c$ and of the time step in the same way that \nbody codes use different time bins for different
particles. The locality of the interactions is an important feature as it makes the parallelisation of the code
straightforward. Running such a simulation on big clusters could thus be easily achieved without having to worry too
much about complex communications and scalability issues.

Let us conclude by stating that our approach has a number of attractive features. Most importantly, the full phase space
information is encoded in the wavefunctions. Working with many wavefunctions, we are in principle able to represent
any given phase space distribution, including those where the velocity dispersion is important. Potentially, this
would allow for numerical simulations of structure formation in presence of warm dark matter.

\section*{Acknowledgements}

This work was supported by the Swiss National Science Foundation and by the
Tomalla Foundation. We would like to thank S.~Cole, A.~Maccio, J.~Read and T.~Theuns, for
useful comments and discussions. O.R. acknowledges the support in part by the National Science Foundation
under Grant No. PHYS-1066293 and the hospitality of the Aspen Center
for Physics.

\appendix

\section{Spherically symmetric case}
\label{sec:spheric}

The framework presented in section \ref{sec:framework} can be simplified in the case of (spatially) spherically
symmetric distribution functions. The dimensionality of the problem is then reduced and allows more comprehensive
convergence studies thanks to the lower number of discretization points needed. 

If we consider only radial motion, then the distribution function can only depend on the distance to the centre $r$,
the radial velocity $v_r$ and the angle between those two vectors. We choose to use the cosine of this
angle as our coordinate, denoted as $y$ in what follows. The gravitational potential does only depend on the distance
to the centre. We thus have $f\equiv f(r,v_r,y)$ and $U\equiv U(r)$. The density at a given $r$ and total mass can be
expressed using these new coordinates and read

\begin{eqnarray}
 \rho(r) &=& \frac{2\pi}{a^3(\tau)} \int_0^\infty v_r^2 dp_r \int_{-1}^1 dyf(r,v_r,y),\\
  M &=& 4\pi \int_0^\infty r^2 dr \rho(r).
\end{eqnarray}
It can be shown that the total mass is a conserved quantity under the equations of motion for $f$. The Vlasov-Poisson
system using those coordinates and assuming spherical symmetry becomes

\begin{eqnarray}
  \frac{\partial f}{\partial\tau} + \frac{yv_r}{a(\tau)}\frac{\partial f}{\partial r} -
a(\tau)\frac{\partial U}{\partial r}\left[y\frac{\partial f}{\partial v_r}+\frac{(1-y^2)}{v_r}\frac{\partial
f}{\partial y}\right] \nonumber\\
+ \frac{(1-y^2)v_r}{ra(\tau)}\frac{\partial f}{\partial y} = 0,
\Bigg.\\
 \frac{1}{r^2}\frac{\partial}{\partial r}\left(r^2 \frac{\partial U}{\partial r}\right)  =
4\pi Ga(\tau)^2 \left( \rho(r) - \bar\rho \right).
\end{eqnarray}
It may, in principle, be possible to find a Wigner-like distribution function for which the Wigner equation corresponds
to
this Vlasov equation. The wavefunctions entering such a distribution would probably obey a spherically symmetric
version of Schr\"odinger's equation. This is, however, not the only way to handle this system. \\
The distribution function can be decomposed in two parts, one for each sign of the coordinate $y$:

\begin{equation}
  f(r,v_r,y) = f_-(r,v_r)\delta_-(y+1) + f_+(r,v_r,\tau)\delta_+(y-1),
\end{equation}
where $\delta_\pm(x)$ are Dirac distributions defined on the interval $[-1,1]$ only. We can then integrate over $y$ and
obtain two equations, one for $f_+$ and another identical up to the signs for $f_-$ together with a boundary condition
ensuring that the two distributions match when they reach $r=0$ or $v_r=0$. The next step in the procedure is to
rescale these distribution functions by introducing $g_\pm(r,v_r) = f_\pm(r,v_r)r^2v_r^2$ and define a combined
distribution $h(r,v_r)$ such that

\begin{equation}
 h(r,v_r) = \left\lbrace
		  \begin{array}{rcl}
                   g_+(|r|,|v_r|) & \rm{if} & rv_r > 0\\
		   g_-(|r|,|v_r|) & \rm{if} & rv_r < 0\\
                  \end{array}
\right.\label{eq:h}
\end{equation}
This new distribution function will obey the following Vlasov equation

\begin{equation}
 \dd{h}{\tau} + \frac{v_r}{a(\tau)} \dd{h}{r} - a(\tau)\dd{U}{r}\dd{h}{v_r} = 0,
\end{equation}
which is identical to the 1D Vlasov equation (\ref{eq:VP}). The difference being in the definition of density and mass
that now read
\begin{eqnarray}
 \rho(r) &=& \frac{2\pi}{r^2R^3(\tau)} \int_{-\infty}^\infty dv_r h(r,v_r, \eta),\\
 M &=& \frac{4\pi^2}{a^3(\tau)} \int_{-\infty}^\infty dr \int_{-\infty}^\infty  h(r,v_r,\eta)db.
\end{eqnarray}
As we are back to the well-known case of Cartesian coordinates (at least for the Vlasov equation), we can introduce the
same decomposition in terms of wave functions than in Section \ref{ssec:WDF}. We will thus solve a set of 1D Cartesian
Schr\"odinger equations alongside a 3D spherically symmetric Poisson equation with a slightly odd density definition.
Using the usual trick $V(r) = U(r)ra(\tau)$, the laplacian term in Poisson's equation can be simplified and the system
we want to evolve reads

\begin{eqnarray}
 i\hbar \dd{\psi_n}{t} &=& - \frac{\hbar^2}{2a(\tau)}\dd{\psi_n^2}{r^2} + m\frac{V}{r}\psi_n
\Bigg.,\\
   \dd{^2V}{r^2} &=& 4\pi Gr\left(
\frac{2\pi}{r^2} \displaystyle\sum_n \lambda_n |\psi_n(r)|^2 - \frac{4\pi^2\Xi}{V_{\rm{tot}}}\right),
\label{eq:spheric_Poisson}
\end{eqnarray}
where $\Xi$ is the normalization of the wavefunctions that can be related to the total mass of the system through

\begin{equation}
  M = \frac{4\pi^2}{a^3(\tau)}\sum_n \lambda_n \int_{-\infty}^{\infty} |\psi_n(r)|^2 dr =
  \frac{4\pi^2\Xi}{a^3(\tau)}.
 \label{eq:normalisation}
\end{equation}
A dynamical term can then be added to equation \ref{eq:spheric_Poisson} to make the framework entirely local as
discussed in Section \ref{ssec:local}. The system can eventually be evolved as if it was a purely one-dimensional
problem. The only difference being the more complicated density terms sourcing Klein-Gordon's (or Poisson's) equation
and the $1/r$ term in the potential of Schr\"odinger's equation.

The generation of initial conditions can be done in exactly the same way than outlined in Section \ref{sec:IC}. The
only difference being the use of the modified distribution $h(r,v_r)$ (equation \ref{eq:h}) instead of $f(r,v_r,y)$ as
the starting point of the procedure.

\nocite{*}
\bibliographystyle{mn2e}
\bibliography{./bibliography.bib}

\label{lastpage}

\end{document}